\documentclass[sigconf, nonacm, prologue, table]{acmart}
\theoremstyle{plain}

\theoremstyle{definition}

\theoremstyle{remark}

\usepackage{multirow}
\usepackage{amsmath}

\AtBeginDocument{%
  \providecommand\BibTeX{{%
    \normalfont B\kern-0.5em{\scshape i\kern-0.25em b}\kern-0.8em\TeX}}}

\begin{document}

\title{A Study on Priming Methods for Graphical Passwords}


\author{Zach Parish}
 \email{zachary.parish@ontariotechu.net}
 \affiliation{%
   \institution{Ontario Tech University}
  \streetaddress{2000 Simcoe Street North, Oshawa, Ontario, Canada}
  \city{Oshawa}
  \state{Ontario}
  \country{Canada}
}

\author{Amirali Salehi-Abari}
\email{amirali.salehi-abari@ontariotechu.ca}
\affiliation{%
  \institution{Ontario Tech University}
  \streetaddress{2000 Simcoe Street North, Oshawa, Ontario, Canada}
  \city{Oshawa}
 \state{Ontario}
  \country{Canada}
}

\author{Julie Thorpe}
\email{julie.thorpe@ontariotechu.ca}
\affiliation{%
  \institution{Ontario Tech University}
  \streetaddress{2000 Simcoe Street North, Oshawa, Ontario, Canada}
  \city{Oshawa}
  \state{Ontario}
  \country{Canada}
}


\begin{abstract}
Recent work suggests that a type of nudge or priming technique called the \emph{presentation effect} may potentially improve the security of PassPoints-style graphical passwords. 
These nudges attempt to prime or non-intrusively bias user password choices (i.e., point selections) by gradually revealing a background image from a particular edge to another edge at password creation time. We conduct a large-scale user study ($n=710$) to develop further insights into the presence of this effect and to perform the first evaluations of its security impacts. We explore the usability impacts of this effect using the subset ($n=100)$ of participants who returned for all three sessions. Our usability analyses indicate that these priming techniques do not harm usability. Our security analyses reveal that the priming techniques can measurably enhance the security of graphical passwords; however, this effect is dependent on the combination of both the image and priming techniques used.
\end{abstract}


\maketitle

\section{Introduction}
User authentication is an integral component for the security of computer systems
, ranging from mobile devices to critical systems.  Its most common form is knowledge-based authentication systems, based on a secret \emph{that you know} (e.g., passwords, PINs, and passphrases) \citep{Herley2012}.
Knowledge-based authentication systems, particularly passwords, are widely popular due to their low cost and lack of specialized hardware requirements.
Unfortunately, password security has become a serious concern, due to recent advances in password attacks using publicly leaked passwords, personal information, and advanced guessing techniques \citep{CCS2016-TARGETEDONLINE, liu19, PERSONAL-PCFG, NNGuessing}. 

These attacks call into question the viability of existing password systems and demonstrate the need for new approaches to improve security. One current approach is to employ complex password-composition policies (i.e. symbol or digit requirements, minimum length requirements, or banning dictionary words), which aim to improve the entropy of the password space. Restrictive policies, however, often lead to user frustration by increasing the difficulty of remembering a secure password \citep{Komanduri2011}. Recently, the National Institute of Standards and Technology (NIST) has recommended against the use of restrictive password composition policies (except password minimum length), arguing that they encourage the creation of minimally complex passwords and/or password reuse \cite{NIST-800-63b}. 
These memorability and security concerns have motivated new authentication methods such as graphical passwords \citep{SUO2005}, where a user remembers an image, or parts of an image, instead of a text password. 

The promise of graphical passwords is to improve password memorability---by using people's superior memory for images---and possibly inspire other ways to improve text passwords \citep{BCV12}. One such graphical password system is PassPoints \citep{wiedenbeck2005}, in which a user is asked to select a sequence of five \emph{click-points} on a background image as a password. The simplicity of PassPoints allows it to serve as a building block or special case of several more complex graphical password systems \citep{MPP,BDAS,CCP,PCCP}.  
Of practical importance is \textit{Microsoft Picture Password} (MPP) \citep{MPP}---an optional login mechanism in Windows 8 and newer versions---for which PassPoints can be viewed as a special case.

PassPoints, among many other graphical password systems, are unfortunately still prone to security concerns where users tend to create predictable graphical passwords, making them easier for attackers to guess \citep{Zhao2015,vanOorschot2010, Thorpe2007, Dirik2007, Sadovnik, Zhu2014}. Successful attacks against PassPoints passwords \citep{vanOorschot2010,Thorpe2007,Dirik2007} have motivated approaches to help users choose unpredictable graphical passwords \citep{PCCP,Bulling2012,katsini2018}. 

One recent approach of interest is a priming technique or nudge that uses an \emph{image presentation} to unobtrusively influence user's graphical password choices at the time of password creation \citep{Thorpe2014}. A special instance of an image presentation is \emph{drawing the curtain}, which slowly reveals a background image, as though a curtain is initially covering it.
A small-scale study ($n=34$) found significant differences in the distribution of first click-points between groups with two different directions of drawing the curtain: right-to-left (RTL) and left-to-right (LTR)  \citep{Thorpe2014}. As the image presentation used in password creation is unknown to an adversary, with a sufficiently large number of image presentation styles for a system, it was suggested to hold promise for complicating guessing attacks, and consequently enhancing security. 

While interesting, the initial image presentation study offered only a suggestion of security improvements; while the study showed a statistically significant impact in user choices, it offered no security analyses. Additionally, it suffers from a number of shortcomings: a small sample size ($n=34$), lack of control group, lack of serious usability analyses, and focusing on only one engineered background image. As such, this paper aims to take a more rigorous approach to evaluating priming techniques by addressing the specific questions:
\begin{enumerate}
    \item How does drawing the curtain affect usability? 
    \item How does drawing the curtain affect security, or password guessability, in practice? 
    \item How do different background images impact the security and usability of drawing the curtain?
\end{enumerate}

We tackle the above questions through a large-scale study of the image presentations ($n=710$), involving three different images (from distinct classes of images) and a control group for each. Our contributions include: 
\begin{enumerate}
    \item  A usability analysis that confirms that the tested priming techniques have no usability impact (measured by SUS scores, login times, password reset rates, and login success rates).
    \item A security analysis that employs a well-known class of automated attacks \citep{vanOorschot2010} in order to better quantify the extent of security improvements that actually arise in practice from such priming techniques. 
    \item An analysis of the priming technique's impact on different background images, which exhibit different saliency distributions \citep{Borji2015}.
\end{enumerate}

Our analyses suggest that these priming techniques tend to improve security.  However, the security improvement is image dependent. For a class of our examined images, the security was considerably improved while for some other classes the security was comparable or slightly improved. Surprisingly, our results indicate that the positive impact of priming techniques on password security is more pronounced on the images with low security potential (e.g, having a single dense saliency region).
Our analyses shed light into the challenges of designing effective priming techniques in graphical passwords, in line with related research for text passwords \citep{Coventry2018, Renaud2017}. 

\section{Related Work}
User authentication methods fall into three categories: what you know (i.e., knowledge-based authentication such as passwords and PINs), what you have (e.g., physical tokens and smartphones), and what you are (i.e., biometrics such as fingerprints).  Knowledge-based authentication offers many benefits including low cost and easy recovery \citep{FRAMEWORK}. Graphical passwords are one knowledge-based proposal that aims to harness the memorability of images.

\vspace{10pt}
\noindent\textbf{Graphical Passwords.}
There is a large body of research on graphical passwords (see these comprehensive surveys \citep{BCV12, SUO2005} for review). Graphical password systems are typically classified into \emph{recall-based} (e.g., BDAS \citep{BDAS}, PassPoints \citep{wiedenbeck2005}, GeoPass \citep{GeoPass_SOUPS,GeoPass_TIFS,addas2019geographical}, etc.) and \emph{recognition-based} (e.g., PassFaces \citep{PassFaces}, VIP \citep{VIP}), depending on whether or not the login phase involves recognizing an image (or set of images).  Recall-based systems are typically further classified into \emph{cued-recall} or \emph{pure-recall}, depending on whether the user is provided an image ``cue" at login time.  

Our focus is a cued-recall system called PassPoints \citep{wiedenbeck2005}, whereby a password is a sequence of five \emph{click-points} on a system-provided background image. A login involves a user selecting the same ordered sequence of five click-points within some acceptable margin of error, referred to as a tolerance region. The simplicity of PassPoints allows it to serve as a building block or special case of several more complex graphical password systems (e.g., MPP \citep{MPP}, BDAS \citep{BDAS}, CCP \citep{CCP}, PCCP \citep{PCCP}). 
One real-world example is \textit{Microsoft Picture Password} (MPP) \citep{MPP}, which serves as an optional login mechanism  in Windows since Windows 8. In MPP, users draw \textit{gestures} (e.g., lines, curves, taps) on a background image to serve as their password. This image is then shown to the user during login to cue password entry. An MPP password created using only the \textit{tap} gesture is analogous to a PassPoints password.

Security issues in PassPoints have been extensively studied, including the consequences of users choosing popular points (or \emph{hot-spots}) \citep{Thorpe2007} and the success of attacks that use image processing \citep{Dirik2007}, human computation \citep{Thorpe2007}, and geometric patterns in the sequence of click-points \citep{salehi2008purely,vanOorschot2010}. In this paper, we apply purely automated attacks \cite{vanOorschot2010} to assess the security of PassPoints passwords. These attacks use a minimal set of click-points that cover the entire image given the error tolerance. The attacks generate sequences of click-points that satisfy the following click-order patterns: \textit{LINE} follows a horizontal or vertical line, \textit{DIAG} follows consistent vertical and horizontal directions (including straight lines in any direction, most arcs, and step patterns), and \textit{LOD} assumes each click-point is within a limited distance from the previous click-point. 

\vspace{10pt}
\noindent\textbf{Background Images.}
Recent work on graphical passwords has explored quantifying the security of background images based on their underlying saliency \citep{Alshehri2016}. In particular, Graph-Based Visual Saliency model \citep{GBVS} with binary thresholding is used to assess the security of background images. Our work extends this approach by placing images into security clusters using their ground truth saliency (as collected by the CAT 2000 dataset \citep{Borji2015}) and considering additional saliency map features alongside regions of interest found through binary thresholding.

\vspace{10pt}
\noindent\textbf{Nudging in Text Passwords.}
Nudging techniques have been applied to traditional text passwords, often in the form of \textit{password meters} that try to nudge users to create stronger passwords by communicating a newly created password's strength. Password meter research has suggested that only stringent password meters (i.e., those that led users to include more digits, symbols and mixed case characters) were able to significantly improve security \cite{ur2012measure}. Transparent \textit{whitebox} password meters have been proposed, whereby a \textit{radar chart} visualization of password security elements (e.g., length, use of digits, use of symbols) is employed to help users improve their passwords 
\cite{hartwig2021nudging}. While users prefer the whitebox password meter over typical (blackbox) password meters, they both offer similar security improvements, even when whitebox meters are personalized based on user's decision making and information processing styles \cite{hartwig2021nudging}.
Password nudges that present a password expiration date (based on the created passwords' strength) was found to significantly improve password security \cite{Renaud2019}. Recently, passphrase memorability was improved by a recognition-based interface and an approach that employed a training period involving semantic priming and a visual implicit learning technique \citep{Joudaki18}. 
 
\vspace{10pt}
\noindent\textbf{Nudging and Priming in Graphical Passwords.}
To improve graphical password security, some nudging techniques have been proposed that limit user's choice during password creation. One approach, Persuasive Cued Click Points (PCCP)\citep{PCCP}, is based on another cued-recall system whereby a user selects a single point on each of five images \citep{CCP}.  PCCP selects a random location to place a viewport, which contains a small region of the image where the user can choose a click-point. Another approach \citep{Bulling2012} uses saliency masks to reduce a user's interest in the most salient parts of the image.

Priming has been explored as a method to help users learn and later recognize a system-assigned ``recognition-based" graphical password \citep{denning2011exploring}. A system called \emph{MooneyAuth} involved priming users with Mooney images to aid in the long-term recollection of a set of system-assigned images and their labels \citep{MOONEYAUTH}. 

Priming has also been explored in user-chosen click-based graphical passwords, using simple image presentations before password creation, and without limiting user choice \citep{Thorpe2014}. The image presentations studied were \emph{drawing the curtain} over a background image to reveal it slowly (from left-to-right or right-to-left), which were found to produce a different distributions in user-chosen click-points. 
Following this work, Katsini et al.\ \citep{katsini2018} devised and studied image presentations that revealed the image starting from the least salient parts, showing the most salient parts last. Other forms of nudging a user towards more secure choices during password creation have also been successfully employed in grid-based graphical passwords \citep{vonZezschwitz2016}

\vskip 10pt
\noindent \textbf{Research Gap.} While the initial study of priming through image presentations (i.e., \textit{drawing the curtain} \citep{Thorpe2014}) suggested that security improvements might result from these priming techniques, it did not provide a concrete security analysis. It also lacked a comprehensive usability analysis to determine if priming techniques negatively impact user experience, which is plausible given that some nudging techniques have been found to annoy users \cite{ur2012measure}. 
A final gap exists in understanding the technique's effect on various background images. In particular, we know that the underlying saliency of an image has an impact on point selection \cite{Thorpe2007, katsini2018}, but the initial study did not provide saliency information for the background image or compare the impact of the technique across images with different saliency distributions. These gaps lead directly to the research questions we seek to answer in this work.

\section{Graphical Password System}
We implemented PassPoints \citep{wiedenbeck2005}, in which a user $u$ is required to select and recall a sequence of 5 click-points (or pixels) on a given background image $I$ as his/her password, denoted by
\begin{equation}
P^{I}_u = \{(x^{I}_{u1},y^{I}_{u1}),\dots,(x^{I}_{u5},y^{I}_{u5})\}.
\label{eq:password_graphical}
\end{equation}
 
Here, $(x^{I}_{ui},y^{I}_{ui})$ is the (x,y)-coordinates that the user $u$ has selected for his/her $i^{th}$ click-points on the image $I$. For enhancing usability, some error tolerance $T$ is permitted for a login attempt's click-points, meaning that each click-point can be within a tolerance distance from each originally selected point. Assuming the login attempt $A= \{(\hat{x}_1,\hat{y}_1),\dots,(\hat{x}_5,\hat{y}_5)\}$, it can successfully login as user $u$ on image $I$ if and only if for all $i \in \{1,\dots, 5\}$:
$$
(|\hat{x}_i - x^{I}_{ui}| \leq T) \text{ and } (|\hat{y}_i - y^{I}_{ui}| \leq T).
$$
In this paper, we set $T=10$ to be comparable with the original study\cite{Thorpe2014} ($T=10$) and other similar studies \cite{chiasson2007,Thorpe2007} ($T=9$). This restriction creates a square 21x21 pixel error tolerance region centered on the selection point.
\subsection{Priming Methods}
During password creation, our system can apply randomly selected priming methods (e.g., \emph{drawing the curtain} \citep{Thorpe2014}) to bias a user's click-points. These priming methods intend to counteract the tendency of users to select similar sequences of click-points when presented with the same background image. Similar password choices form \emph{hot spots} \citep{Thorpe2007} and \emph{click-order patterns} \citep{vanOorschot2010} which undermines the theoretical security promises of PassPoints.  

We implement the \emph{drawing the curtain} priming method \citep{Thorpe2014}, where users watch the target image being gradually revealed before selecting their click-points. In our system, the user is first presented with a blank white image, and the target image is gradually revealed starting from one edge over 20 seconds.  In replicating the drawing-the-curtain method, we apply curtain drawing in the left-to-right (LTR) and right-to-left (RTL) directions. We chose to study RTL and LTR exclusively, rather than alternative methods such as top-to-bottom or bottom-to-top, so that our work could be based on a known result before being extended to other possible priming techniques.

\section{Image Selection Methodology}\label{sec:img_sel}
\begin{figure}
    \centering
    \includegraphics[width=0.46\textwidth]{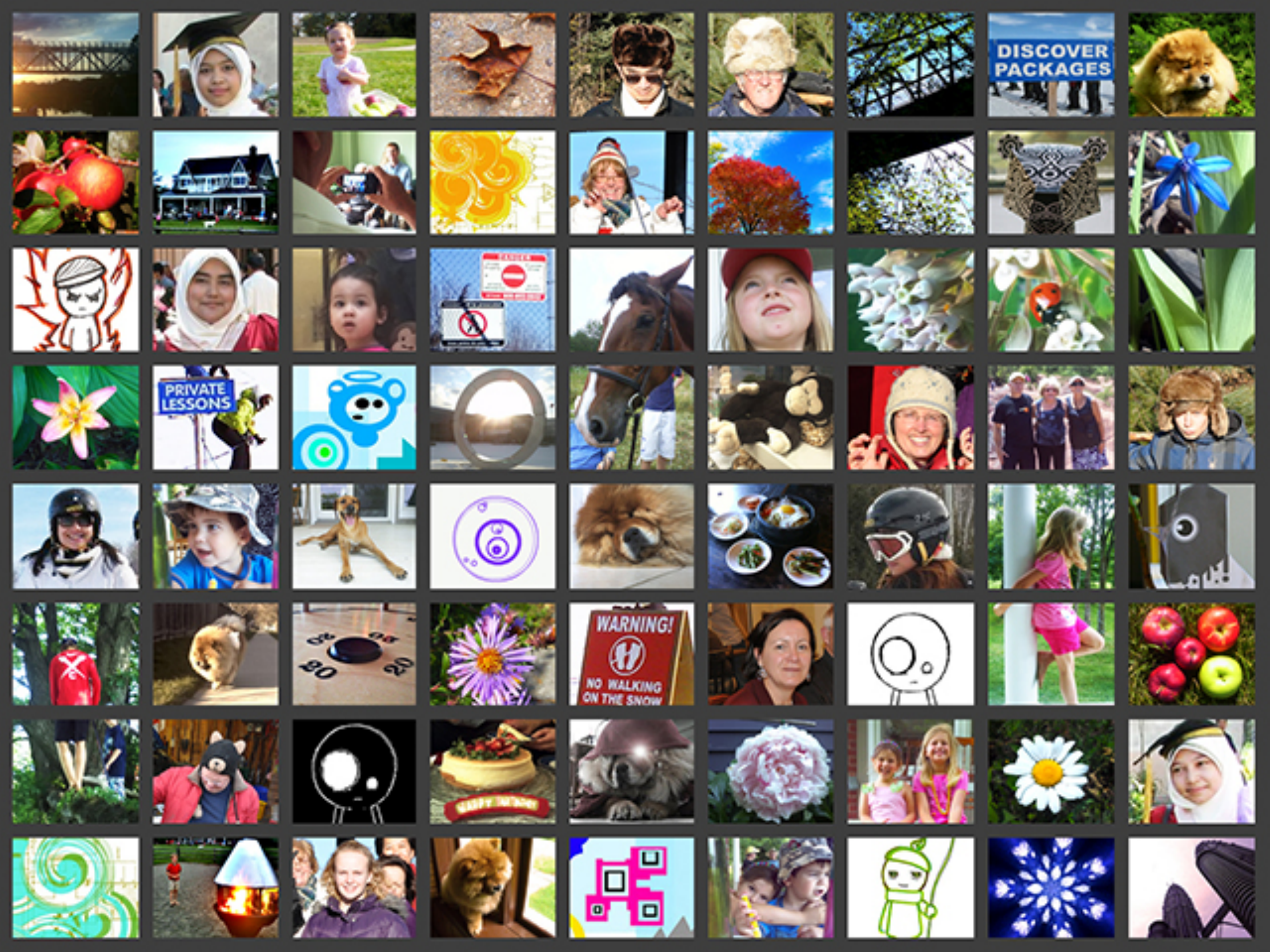}
    \caption{Grid image \citep{Thorpe2014}}
    \label{fig:grid_img}
\end{figure}

The presentation effect was previously examined on a grid image composed of smaller images (see Figure \ref{fig:grid_img}) \citep{Thorpe2014}. We investigate the drawing-the-curtain technique on this grid image, and two additional non-composite images, carefully selected to present different underlying saliency with more natural real-life stimuli. We select these images from the CAT2000 dataset \citep{Borji2015} which contains thousands of images and their ground truth saliency maps. We preprocess the dataset and extract features from the saliency maps to perform clustering. Our clustering aims to identify distinct classes of images with structurally different distributions of saliency regions. We hypothesize that the manner in which saliency is distributed over an image will play a role in the security of passwords created on the image, and have an impact on the effect of priming techniques. In particular, we expect that images where saliency is less evenly distributed will produce more predictable passwords, and thus be more susceptible to guessing attacks.  We find three distinct clusters and select the center image of each cluster as its representative. For comparability with the grid image, the selected images are scaled to a 680x460 resolution. We explain the details of this selection process below.

\subsection{CAT2000 Dataset}
The CAT2000 dataset \citep{Borji2015} contains 2000 images from 20 categories (e.g., Indoor, Outdoor Natural, Object, etc.) with eye-tracking fixation point data generated by 18 observers performing a five-second free look on each image. For each image, a greyscale saliency map is generated by smoothing the fixation points of all viewers to approximate the continuous distribution generated by infinite viewers. Each pixel in the saliency map ranges from 0 to 255.  The higher a pixel value is, the higher its saliency is. A pixel with a value of 0 indicates that the pixel was not salient to the observers. The distribution of saliency within each map varies greatly, ranging from maps with a single small region of high-intensity saliency to maps with saliency distributed more uniformly across the image in large regions.

\subsection{Preprocessing}
We focus on seven image categories of the CAT2000 dataset: Action, Indoor, Object, OutdoorManMade, OutdoorNatural, Random and Social. We have excluded categories composed of art or computer generated graphics (e.g., Sketch, Cartoon, LineDrawing) and categories with a particular visual effect (e.g., noisy, jumbled, inverted). This exclusion allows us to focus on images  primarily drawn from real life scenes with minimal artificial visual stimuli types. We also exclude the \textit{Affective} category to remove potentially disturbing imagery for the subjects in our study.

As with the original implementation of the presentation effect, our system uses 640x480 background images. Since we must resize our selected images to fit this size, we consider only images with a 4:3 aspect ratio to prevent distortion.\footnote{For eye-tracking purposes all images in the CAT2000 dataset were superimposed onto a 1920x1080 grey background image, creating grey image borders. We remove these borders from the images and from their associated saliency map.} From this set of images we select only those with a resolution of 1440x1080 so that image sizes are the same for clustering.

\subsection{Feature Extraction}
We next extract six features from the saliency maps of the candidate filtered images to construct \emph{image feature vectors}. Our image feature vectors are specifically designed to capture the number, spread, and density of saliency regions within each image. For each saliency map, our features include: \textit{(i) Salient Proportion}: fraction of non-zero valued pixels; \textit{(ii) All Pixel Variance}: variance of all pixels; \textit{(iii) Salient Pixel Variance}: variance of all non-zero valued pixels; \textit{(iv) Number of Saliency Regions}: number of unconnected regions of salient pixels when we threshold the saliency map using Otsu's method \citep{Otsu1979}; \textit{(v) Distance Between Saliency Regions}: distance between unconnected regions of salient pixels after Otsu's thresholding; and \textit{(vi) Proportion of High Saliency Pixels}: fraction of non-zero valued pixels after Otsu's thresholding. These features were selected from a larger set of candidates as they provided the most reasonable clustering result during manual inspection. For this process, we first observed that saliency maps can usually fall into two categories. The \emph{compact saliency maps} are concentrated in single or few points whereas the \emph{diffuse saliency maps} have a more even distribution of saliency points. From the pool of images, we manually selected two small subsets of representatives of compact and diffuse saliency distributions. We then clustered all images using various features. The final selected features (as reported above) were the features that resulted in clusters consistent with the diffuse and compact categories.  

\subsection{Clustering}
We cluster the image feature vectors using the k-means algorithm \citep{macqueen1967} (initialized with k-means++ \citep{Arthur2007}) with k=3 as determined by the Kneedle algorithm \citep{Satopaa2011}. This yields three clusters that are largely stable across different random initializations of k-means. For each cluster, we select an image with the shortest distance to its cluster's centroid as the representative for that cluster. As k-means is non-deterministic, mainly due to random initialization, the representative image might sometimes be different for different initializations. We select the representatives that were the most common in 1000 different run of k-means with random initializations. In our case, all three cluster representatives were consistently selected in all 1000 runs.

\begin{figure*}
    \begin{center}
        \begin{tabular}{ccc}
            \includegraphics[width=0.3\textwidth]{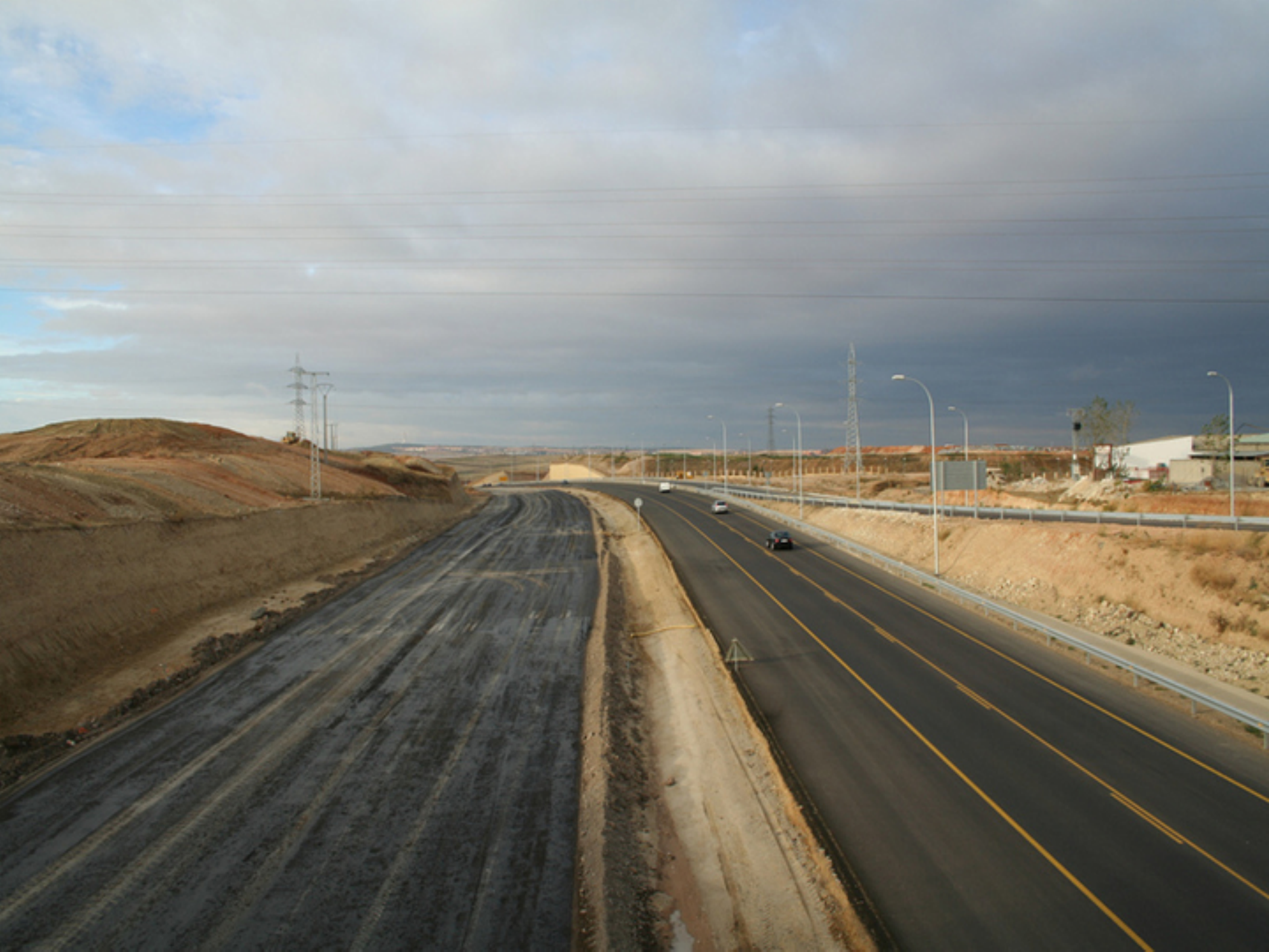} 
            & \includegraphics[width=0.3\textwidth]{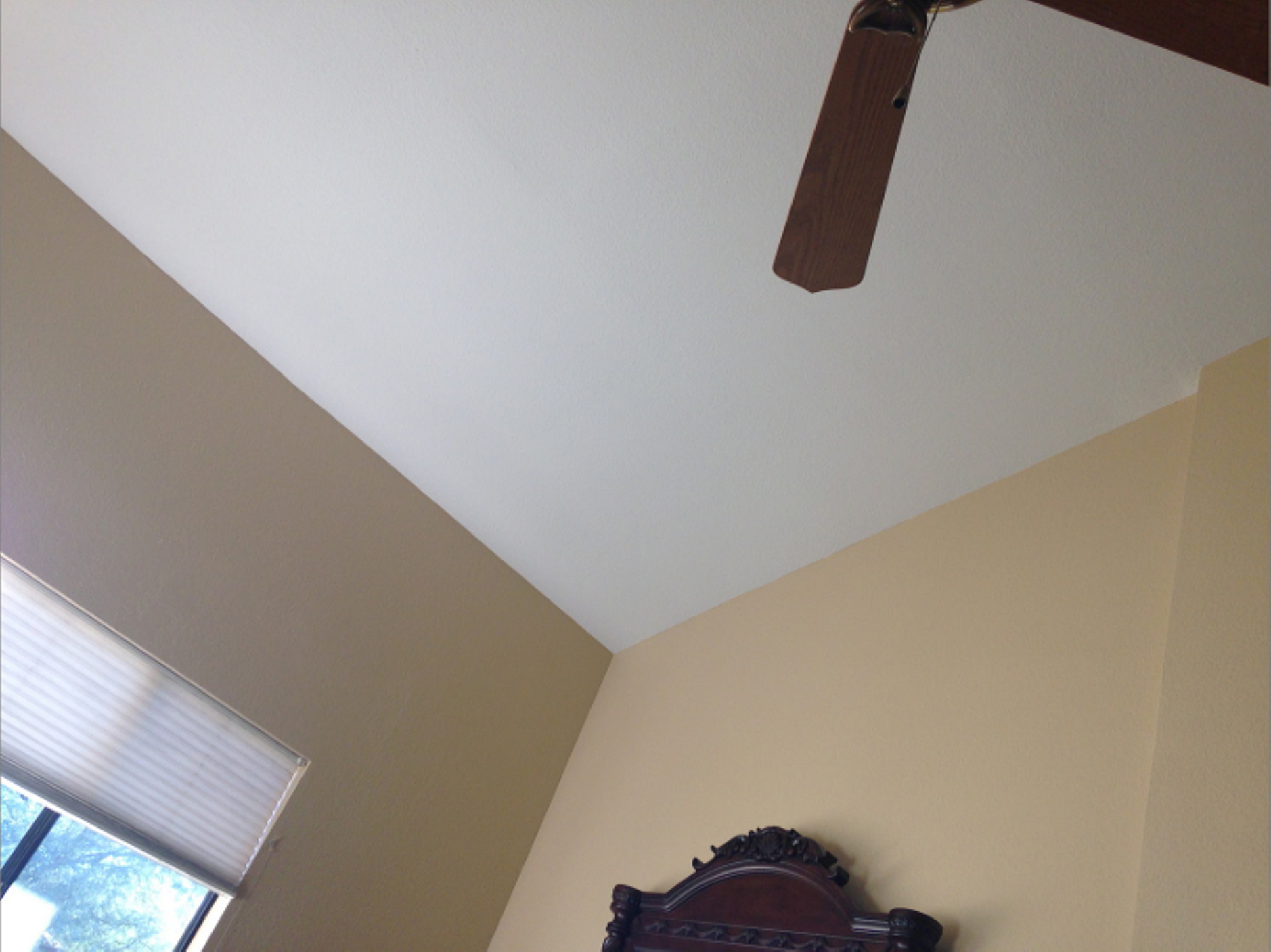} 
            & \includegraphics[width=0.3\textwidth]{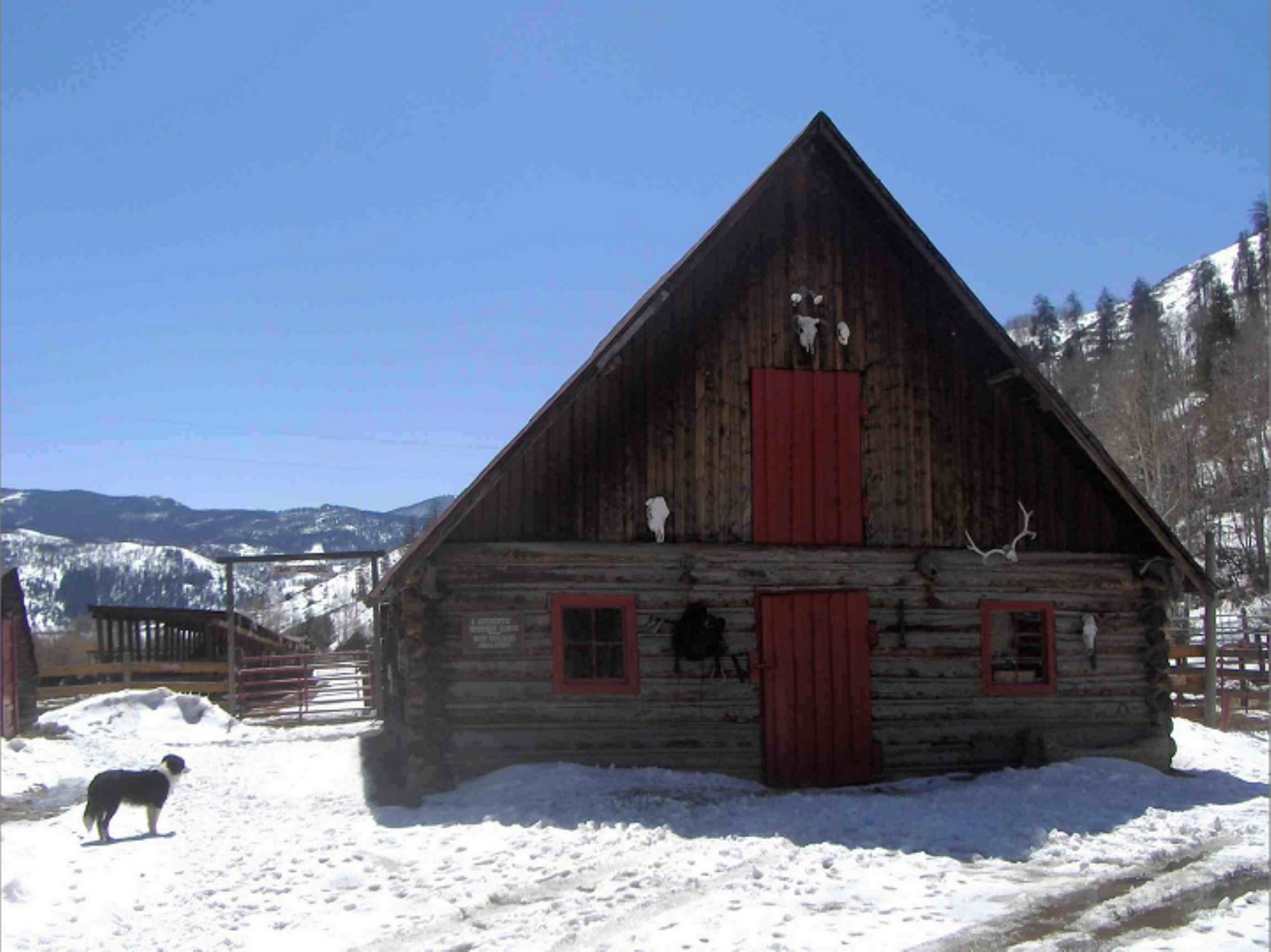} \\
            (a) Highway Image & (b) Fan Image & (c) Barn Image\\\\
            \includegraphics[width=0.3\textwidth]{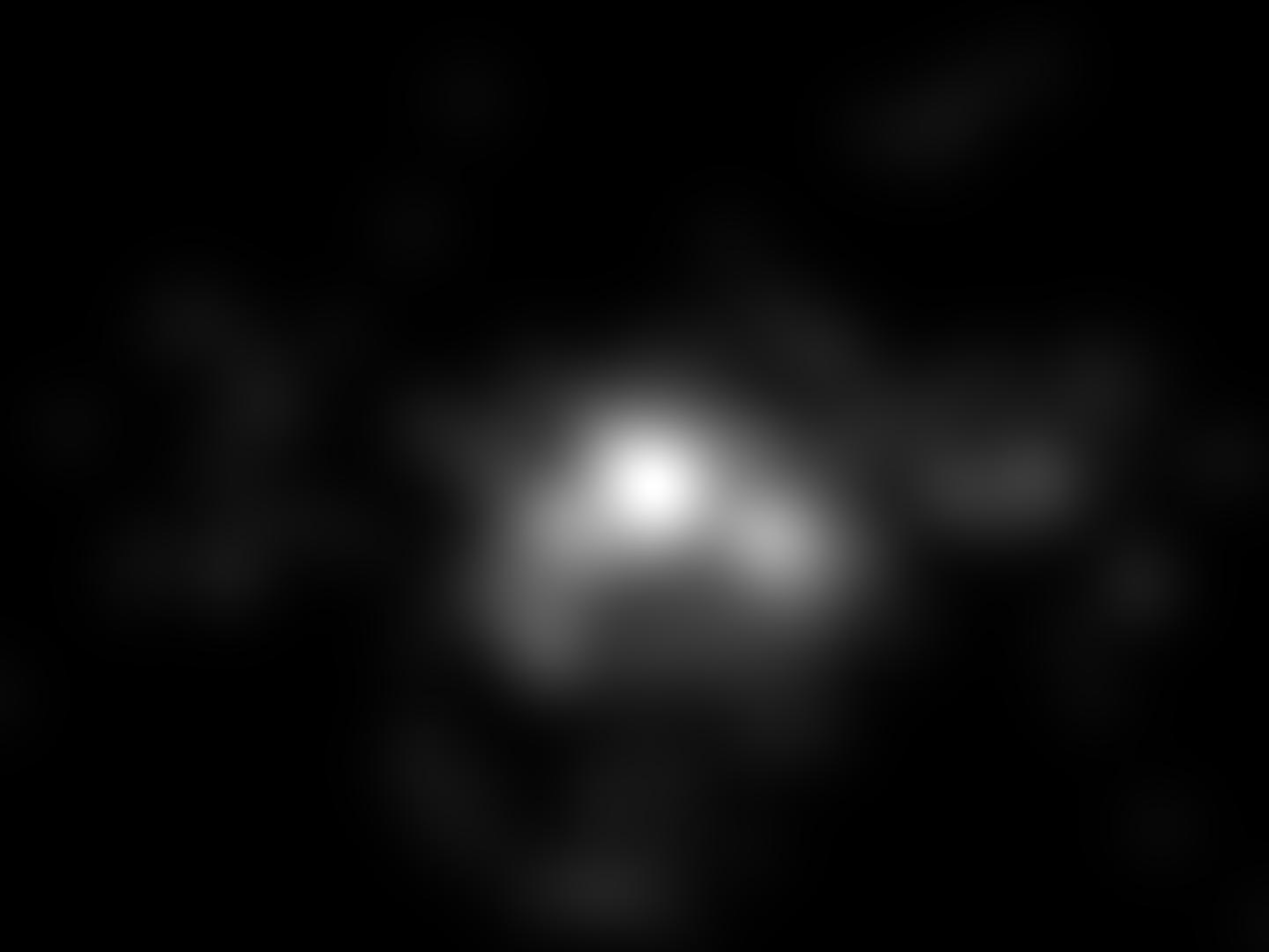} & \includegraphics[width=0.3\textwidth]{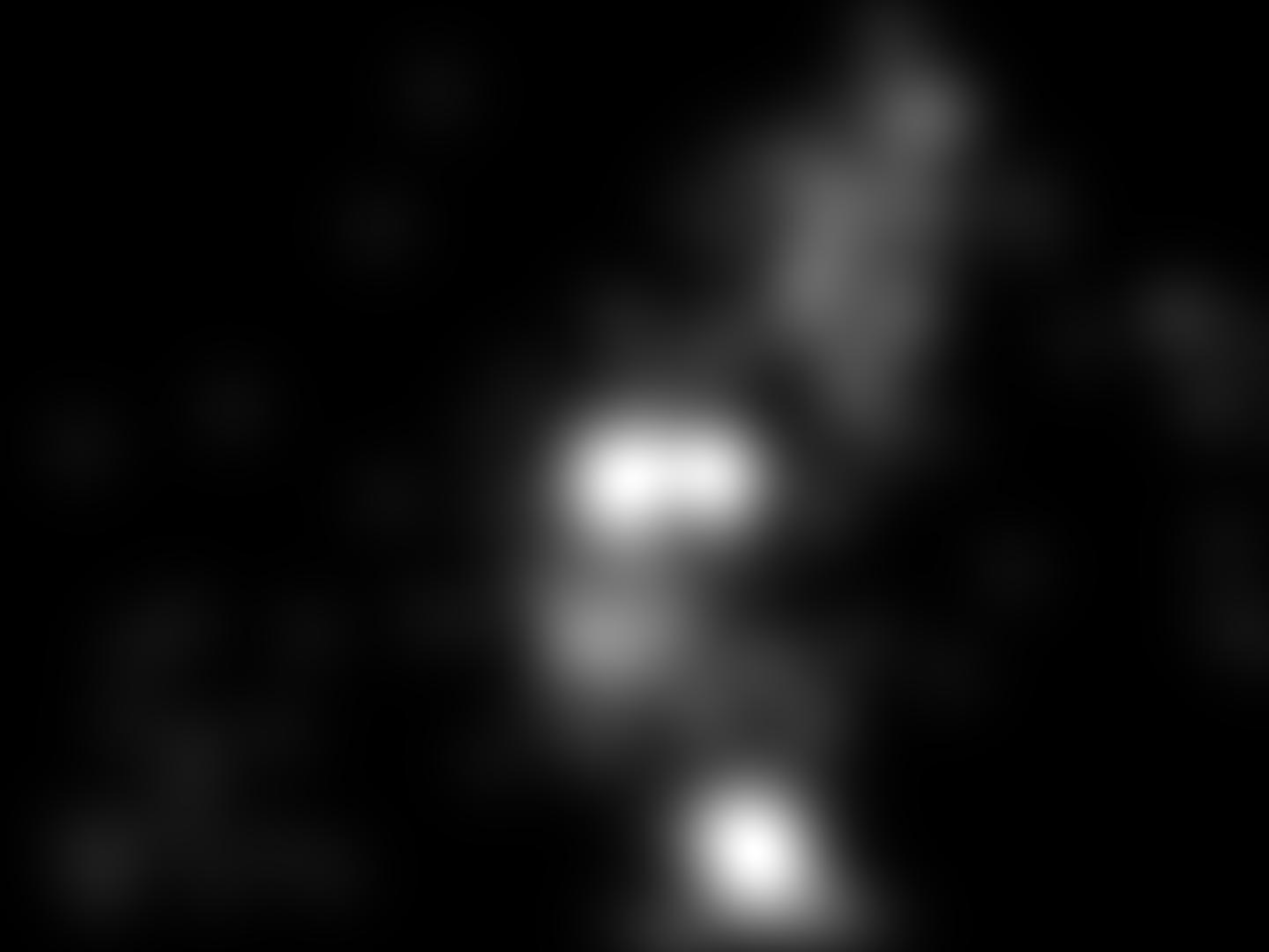} & \includegraphics[width=0.3\textwidth]{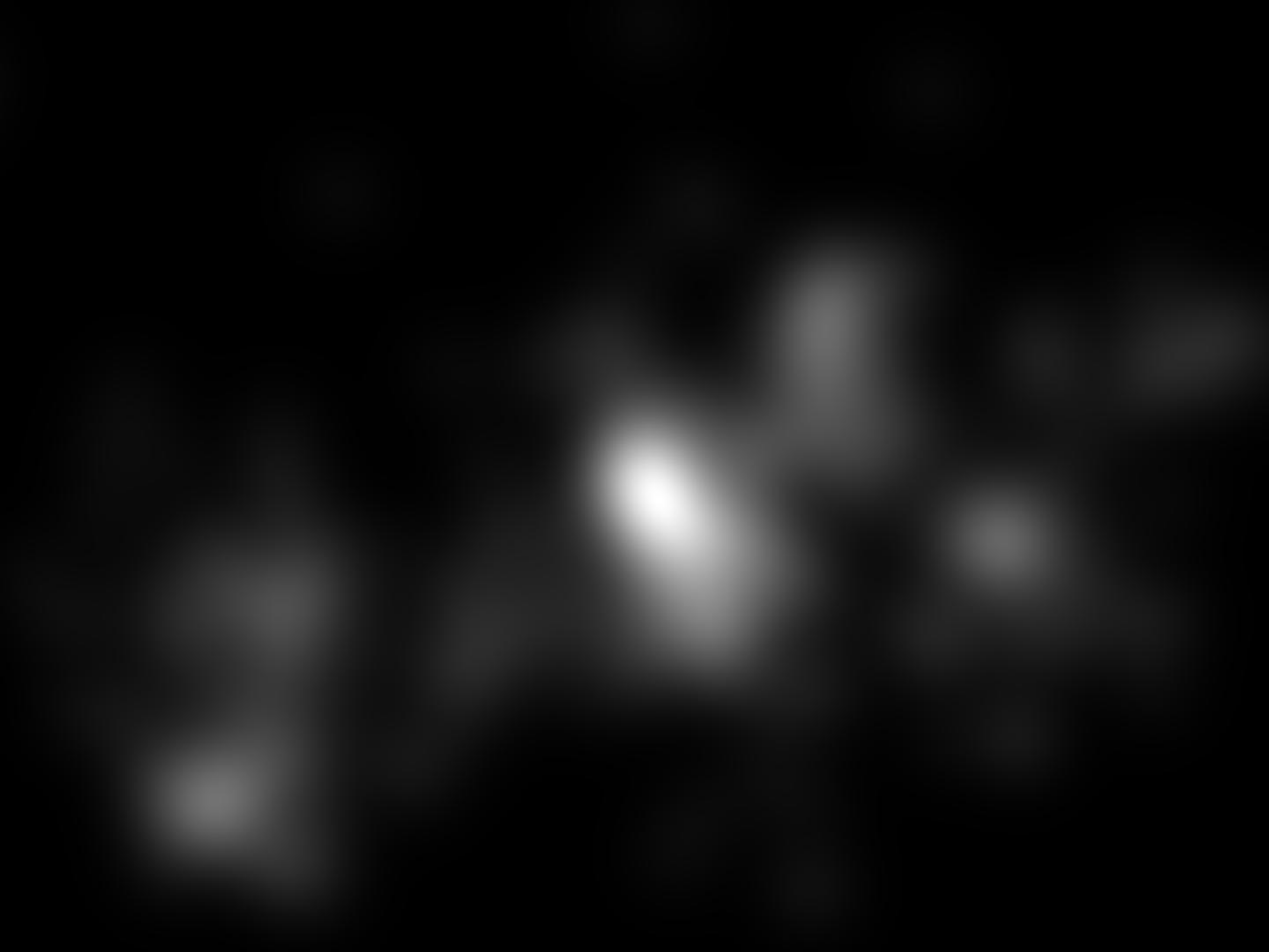} \\
            (d) Highway Image Saliency Map & (e)
            Fan Image Saliency Map & (f) Barn Image Saliency Map \\
        \end{tabular}
    \end{center}
    \caption{Selected background image (top) and its associated saliency map (bottom). Each image is a representative of its associated cluster: Highway (compact cluster), Fan (outside cluster) and Barn (diffuse cluster).}
\label{fig:bkg_img_sal}
\end{figure*}

The three detected clusters exhibit relatively distinct characteristics of saliency region distributions. The \textit{compact} cluster contains images that have a small, dense, and typically center-biased, region of salient visual information within a largely non-salient image (see Figure \ref{fig:bkg_img_sal}(d) for an example). The Highway image in Figure \ref{fig:bkg_img_sal}(a) is the selected representative of this cluster. The \textit{diffuse} cluster contains images where saliency is spread more uniformly throughout the image in moderately dense saliency regions. The Barn image in Figure \ref{fig:bkg_img_sal}(c) is this cluster representative. The \textit{outside} cluster contains images that fall outside of the other two clusters. Typically images in this cluster have several saliency regions that are too large or too distant from each other to be comparable to the compact images, but not large or widely spread enough to be clustered with the diffuse images. This cluster's representative is the Fan image in Figure \ref{fig:bkg_img_sal}(b). As images in the outside cluster typically present a blend between characteristics found in the compact and diffuse clusters, we ignore this cluster and perform our user studies using only the Highway and Barn images. We expect that images in the diffuse cluster will present more possible click-point locations to the user, leading to more entropy in the password space, and therefore higher resiliency to guessing attacks.

\section{User Study}
Our user study is split into 3 sessions across 8 days. Users are directed to interact with a graphical password authentication scheme deployed on our website. They create a graphical password, and log in with it several times over 8 days to simulate average reported time between logins for email and online banking platforms \citep{Hayashi2011}. We ask users to complete a questionnaire to record demographic and usability information. This user study, questionnaire, and an exit survey were approved by our institution's research ethics board. Our study employs a between-subject design where we compare users exposed to a priming effect to users in a control group. We detail how each session proceeds, our recruitment, and demographics below. 
\subsection{Sessions and Procedures}
For each selected image, we run a separate user study with sessions and procedures discussed below.
\vskip 3mm
\noindent \textbf{Session 1 (day 1).}
Users are recruited from the crowd-sourcing website Amazon Mechanical Turk (MTurk) and directed to visit our website. Users are told they will be participating in a usability test of an graphical password system, but are not told about the priming. Users visiting the website with mobile devices are automatically detected and filtered out. Mobile devices are excluded from the study in order to normalize user screen sizes and the mode of interaction with the system. Our 640x480 images are sized to fit cleanly, without distortion or scrolling on all but the smallest screens. Users then enter their MTurk ID number as a username and are randomly assigned upon enrollment to one of three groups: left-to-right (LTR), right-to-left (RTL), or control. Users in the LTR and RTL groups are primed by a \emph{drawing the curtain} image presentation that begins revealing the image from the left side and the right side, respectively. Users in the control group are not primed. The groups were approximately equal in size. However, after filtering out disqualified users (see Section \ref{sec:filtering}), the control group is larger than LTR and RTL groups (see Table \ref{tab:number_users}).


Users then watch a short instruction video detailing how to create and login using our graphical password system. Next, they create a practice password on a practice background image to familiarize themselves with the system. For consistency, this practice background image is revealed using the same type of curtain drawing method (e.g., LTR, RTL, or none) that they will be exposed to when they create their real password. Users are then instructed to select a password to login with for this and other sessions on a specific background image primed according to their group.

Users then fill out a demographics questionnaire on their sex, age, first language, field of work or study, level of experience with computers, and level of experience with computer security. Users are then asked if they have seen their password's background image prior to the study, and if they used a touch screen device for the study. Users are then prompted with their background image and asked to click the first object that drew their attention. For users in a treatment group with a priming effect, we ask if they watched the entire curtain draw, or were distracted during the effect. Users then select from a drop down menu their strategy for creating their password, and are asked to provide further details in text. Users can select from the following strategies: (i) I used colours for selecting my graphical password; (ii) I used shapes for selecting my graphical password; (iii) I used geometric patterns (e.g., lines, circles, corners, etc.) for selecting my graphical password; (iv) I used the first object that drew my attention for selecting my graphical password; and (v) Other. 

Users are then shown their background image again and asked to login with their selected password. If users cannot successfully login, they can reset their password and create a new one on the same background image with the same priming effect. Upon a successful login, this session ends.

\vskip 3mm
\noindent \textbf{Session 2 (day 2--3).}
Session 2 takes place 24--48 hours after Session 1 for each user, in order to simulate self-reported frequency of logging into email accounts \citep{Hayashi2011}. Users were notified about the second session through the MTurk platform 24 hours after completing Session 1.

Users are directed to our website and start by entering their ID. After watching a video that instructs them to login with their password from Session 1, they  are prompted to login and shown their background image. Users who have forgotten their passwords can reset from this session to Session 1 to create a new password. They must then wait an additional 24 hours to attempt Session 2 a second time. Users who reset their password are placed in the same priming group and shown their same original background image. After users successfully login with their password, Session 2 ends.
\vskip 3mm
\noindent \textbf{Session 3 (day 7-8).}
Session 3 takes place five days after Session 2 to simulate average reported time between logins for online banking platforms \citep{Hayashi2011}. Users were notified about this session through  MTurk 5 days after completing Session 1. Users first enter their ID, then watch a short video explaining the session, and are then prompted to login using their background image. After logging in successfully, users fill out an exit survey. The survey asks users if they used a touch screen for any sessions and if they recorded their password externally during the study. If they were in a group with a priming effect, they are also asked whether they noticed the effect was only used during creation. The exit survey also includes a System Usability Scale (SUS) survey to collect usability information about the system \citep{brooke1996}. Users who cannot successfully log in during this session are not allowed to reset and are given the exit survey to complete.

\subsection{Participant Recruitment}
All participants in our studies are recruited through Amazon's Mechanical Turk platform. An advertisement was listed for the first session and users were notified of the subsequent sessions through the platform as they became available. Users could only complete Sessions 2 and 3 if they had completed the previous sessions. Users were paid \$1.50 USD for Session 1, \$0.25 USD for Session 2 and \$0.75 USD for Session 3.  Throughout the advertisements, consent forms, and system itself, participants are told they will be participating in a usability test of a graphical password system, but are not told about the priming. Our study and compensation structure was approved by the ethics board of our institute. 

For the Grid image, Sessions 1-3 are run with 436, 188, and 124 participants respectively. For the Highway and Barn images, we conduct only Session 1, with 216 and 213 participants respectively, as our primary goal was security analyses for those images. In total, 865 participants completed Session 1, but after filtering the data as described in Section \ref{sec:filtering}, 710 participants remain for our analysis. Table \ref{tab:number_users} reports the final number of users for each group and image in Session 1. 

\begin{table}[tb]
    \centering
    \setlength{\tabcolsep}{10pt}
    \begin{tabular}{lccc}
         \toprule
         &\textbf{Control} & \textbf{RTL} & \textbf{LTR} \\
         \midrule
         \textbf{Grid}& 157 & 95 & 125 \\
         \textbf{Barn}& 79 & 37 & 38 \\
         \textbf{Highway}&70 & 50 & 59 \\
         \bottomrule
    \end{tabular}
    \caption{The number of users in each group of user study after filtering out disqualified users for Session 1.}
    \label{tab:number_users}
\end{table}

\subsection{Demographics}
Here we detail the self-reported demographic data collected in Session 1 for each image. This includes only qualified users (see Section \ref{sec:filtering} for more information on the filtering process). For the Grid image, we had 377 participants with 161 completing Session 2 and 107 completing Session 3. The self-declared gender identification was 252 (66.8\%) male vs 125 (33.2\%) female. Some other notable demographic information is that 347 (92.0\%) of participants were first-language English speakers, 80 (21.2\%) were students, and 243 (64.5\%) were 35 years or younger.  For computer and security skills, 366 (97.0\%) of participants self-reported their computer skills to be 3 or above and 306 (81.2\%) of participants reported their computer security skills as 3 or above. We found similar demographic breakdowns for the Highway and Barn images (see Table \ref{tab:demographics}), with the exception that the Barn group contained more students. We found no change to our results when we performed comparisons across demographic segments.

\begin{table*}
    \begin{center}
        \begin{tabular}{|l||l|l||l|l||l|l||l|l||l|}
         \hline
          &\multicolumn{8}{c||}{\textbf{Participant Demographics}}\\
         \hline
         \hline
         \multirow{6}{*}{\textbf{Grid}} &\multicolumn{2}{ c||}{\textit{Age}} &\multicolumn{2}{ c||}{\textit{Gender}} &\multicolumn{2}{ c||}{\textit{Computer Skills}} &\multicolumn{2}{ c||}{\textit{Occupation}}\\
         \cline{2-9}
         &$<$20 &26 (6.9\%) &Male &252 (66.8\%) &1-2 &11 (3.0\%) &Student &80 (21.2\%)\\
         \cline{2-9}
         &20-25 &43 (11.4\%) &Female &125 (33.2\%) &3-5 &366 (97.0\%) &Non-Student &297 (78.8\%)\\
         \cline{2-9}
         &25-30 &96 (25.5\%) &\multicolumn{2}{ c||}{\textit{Language}} &\multicolumn{2}{ c||}{\textit{Security Skills}} &\multicolumn{2}{ c||}{\textit{Work/Study Major}}\\
         \cline{2-9}
         &30-35 &78 (20.7\%) &EN &347 (92.0\%) &1-2 &71 (18.8\%) &CS/IT &93 (24.7\%)\\
         \cline{2-9}
         &$>$35 &134 (35.5\%) &OTH &30 (8.0\%) &3-5 &306 (81.2\%) &OTH &284 (75.3\%)\\
         \hline
         \hline
         \multirow{ 6}{*}{\textbf{Highway}} &\multicolumn{2}{ c||}{\textit{Age}} &\multicolumn{2}{ c||}{\textit{Gender}} &\multicolumn{2}{ c||}{\textit{Computer Skills}} &\multicolumn{2}{ c||}{\textit{Occupation}}\\
         \cline{2-9}
         &$<$20 &18 (10.0\%) &Male &119 (66.5\%) &1-2 &6 (3.4\%) &Student &41 (22.9\%)\\
         \cline{2-9}
         &20-25 &28 (15.6\%) &Female &60 (33.5\%) &3-5 &173 (97.7\%) &Non-Student &138 (77.1\%)\\
         \cline{2-9}
         &25-30 &43 (24.0\%) &\multicolumn{2}{ c||}{\textit{Language}} &\multicolumn{2}{ c||}{\textit{Security Skills}} &\multicolumn{2}{ c||}{\textit{Work/Study Major}}\\
         \cline{2-9}
         &30-35 &29 (16.2\%) &EN &164 (91.6\%) &1-2 &29 (16.2\%) &CS/IT &47 (26.3\%)\\
         \cline{2-9}
         &$>$35 &61 (34.1\%) &OTH &15 (8.4\%) &3-5 &150 (83.8\%) &OTH &132 (73.7\%)\\
         \hline
         \hline
         \multirow{ 6}{*}{\textbf{Barn}} &\multicolumn{2}{ c||}{\textit{Age}} &\multicolumn{2}{ c||}{\textit{Gender}} &\multicolumn{2}{ c||}{\textit{Computer Skills}} &\multicolumn{2}{ c||}{\textit{Occupation}}\\
         \cline{2-9}
         &$<$20 &4 (2.6\%) &Male &89 (57.8\%) &1-2 &7 (4.5\%) &Student &52 (33.8\%)\\
         \cline{2-9}
         &20-25 &26 (16.9\%) &Female &65 (42.2\%) &3-5 &147 (95.5\%) &Non-Student &102 (66.2\%)\\
         \cline{2-9}
         &25-30 &46 (29.9\%) &\multicolumn{2}{ c||}{\textit{Language}} &\multicolumn{2}{ c||}{\textit{Security Skills}} &\multicolumn{2}{ c||}{\textit{Work/Study Major}}\\
         \cline{2-9}
         &30-35 &32 (29.8\%) &EN &132 (85.7\%) &1-2 &22 (14.3\%) &CS/IT &34 (22.1\%)\\
         \cline{2-9}
         &$>$35 &46 (29.9\%) &OTH &22 (14.3\%) &3-5 &132 (85.7\%) &OTH &120 (77.9\%)\\
         \hline
         \end{tabular}
    \end{center}
    \caption{Demographic information for all studies over three background images of Grid, Highway, and Barn.}
    \label{tab:demographics}
\end{table*}

\subsection{Filtering and Outliers}
\label{sec:filtering}
One challenge of using MTurk is that participants might have low incentive to perform as well as a natural setting. To reduce the impact of such participants, we use some responses to the exit survey of the first session to inform a filter. For RTL and LTR groups's exit survey, the users were asked to report if they watched the image reveal and if they were distracted from watching the reveal (e.g., by switching to another browser tab during the reveal or focusing their attention away from their screen). Users who responded that they did not watch the reveal, or reported being distracted during the reveal are filtered out from our analysis. 

During analysis, we discovered one notable group of outliers who selected their click-points in a different manner than most participants. These users selected their points repeatedly in the same, or very close (i.e., within the error tolerance region) to the same location. We include these users in our analysis as we find it likely that this behavior would be present in a real-world application of this system, similar to the poor password creation behaviors observed in text passwords.

\section{Results}
We report the results of our user studies with regard to point selection biasing, usability analyses, and security analyses.

\subsection{Selection Point Biasing}
We first attempt to replicate the statistical findings of the presentation effect demonstration \citep{Thorpe2014}. We examined whether the x-coordinates for each of the five click-points come from the same distribution, when RTL and LTR presentation groups are compared. 

We extend the original paper's tests by including a control group whose members create their passwords without any priming effect. For each background image, we compare the passwords generated by each presentation treatment group with the passwords generated by that image's control group, yielding two sets of tests for each image: RTL vs. Control and LTR vs. Control. 
We let $x^I_{ui}$ represent the x-coordinate of the $i^{th}$ click-point selected by user $u$ on image $I$. Given this notation, and to be consistent with the original paper \citep{Thorpe2014}, we formulate a class of  null hypotheses of the form:

$\mathcal{H}^{(I,G,i)}_{0}$: the two samples $\{x^{I}_{ui} | u \in G\}$ and $\{x^{I}_{ui} | u \in C\}$ come from the same distribution.

Where $C$ refers to the control group and $G$ can be any priming treatment group of LTR and RTL. From the previous work \citep{Thorpe2014}, one expects that users' first click-points (i.e., $i=1$) will be biased towards the edge that the curtain drawing began from (e.g., right for RTL group and left for LTR). We therefore test 5 null hypotheses for each primed treatment group against a control group for each image. For these tests we compare the click-points of all users who completed Session 1 of the study on a particular image. In order to be comparable with the results of the original paper, we begin by testing each of these hypotheses using a one-sided Mann-Whitney-U test (with $\alpha$ = 0.05).

\begin{figure}
    \centering
    \includegraphics[width=0.4\textwidth]{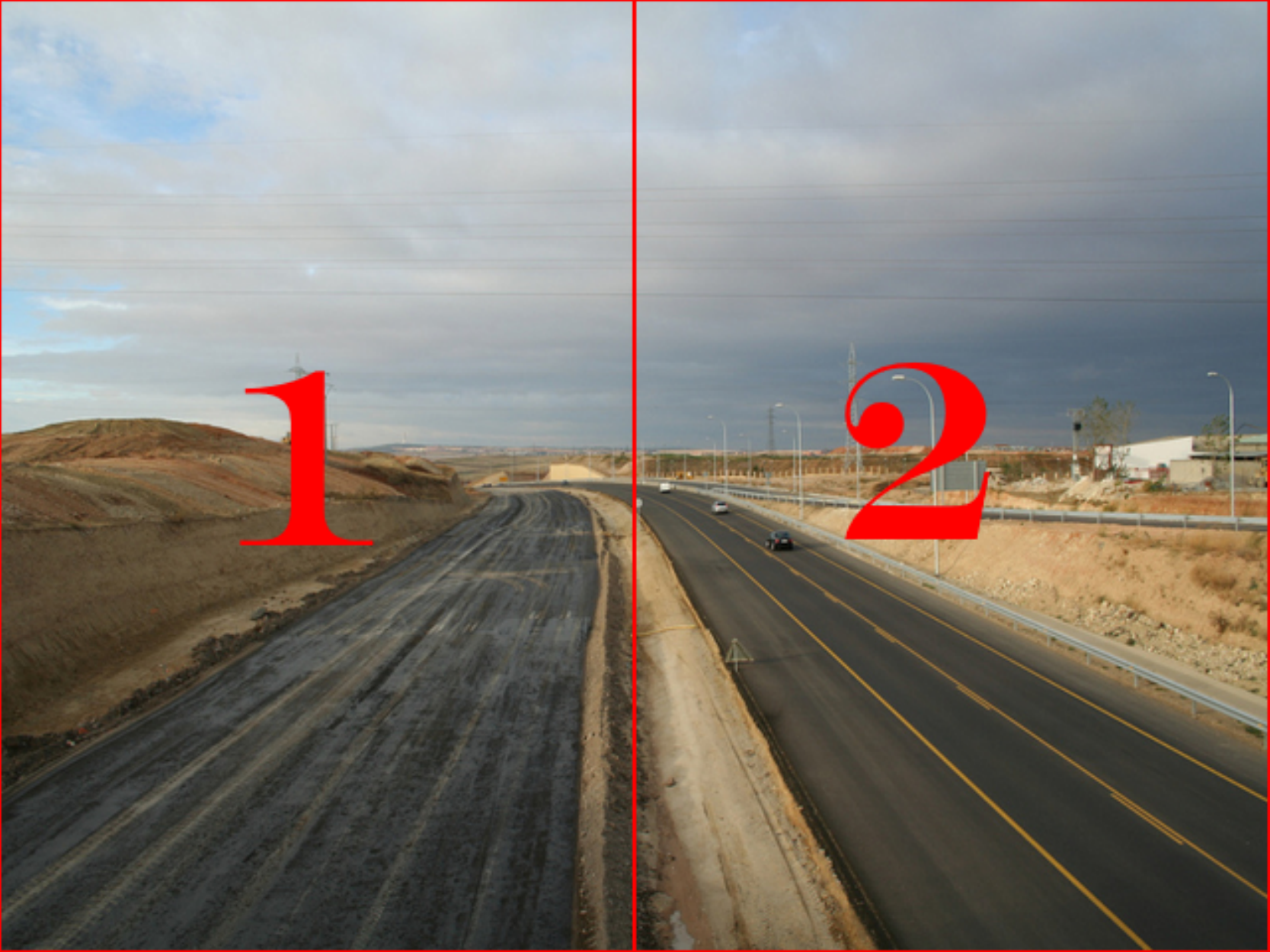}
    \caption{Each background image is divided into two equally sized bins for the Fisher exact test.}
    \label{fig:bins}
\end{figure}

We first replicated the RTL vs. LTR test performed in other work \citep{Thorpe2014}, which suggested that the x-coordinates for the first click-points of two opposing groups (i.e., RTL and LTR) were statistically different ($p=0.019$ without multiple testing correction), but which was insignificant after a Bonferroni correction for the five tests ($p=0.091$). Similar to the other work, we find an insignificant result on the first click-points between RTL and LTR after a Bonferroni correction for the five tests ($p = 0.1156$, effect size = 0.13).

For both the RTL vs Control and LTR vs Control test pairs on all three images (i.e., Grid, Highway and Barn), we fail to reject any of our null hypotheses with the Mann-Whitney-U test, suggesting that the presentation effect does not bias the x-coordinates of click-points in a statistically significant way for these images. All p-values were insignificant even before correcting for multiple tests. 

To determine if the priming effects have an impact at a coarser level of granularity, we also test each hypothesis with a Fisher exact test. For these tests, we divide each image into two equally sized bins that span the height of the image and half of the width (see Fig. \ref{fig:bins}), and record the distribution of points over the bins for the RTL, LTR and Control groups, for each click-point.


When applying the Fisher exact test to our hypotheses, we fail to reject any null hypotheses for either treatment group vs.\ Control for any of our images after performing a Bonferroni correction. 
This suggests that priming did not impact the distribution of click-points for these images even on a coarse level.

While both statistical tests can capture a change in distribution along the x axis of the image, neither capture the impact of priming on the security of the generated passwords. It is possible that the priming effects alter the security of the generated passwords, without changing the distribution of x values in a way that is significant. This has motivated our security analyses below.

\subsection{Security Analysis}
\label{sec:security}
We test the security of each group's passwords against three classes of well-known purely automated click-order based attacks \citep{vanOorschot2010}. Each attack first creates an alphabet based on the background image's resolution and error tolerance $T$. This alphabet is constructed by tiling the background images with squares of $(2T+1)\times(2T+1)$ pixels.\footnote{For the edges of the image, we allow the squares to overhang the edge.} The (x,y)-coordinate of each square's center corresponds to an element in the alphabet.\footnote{As our images have the resolution of 640x480 and the error tolerance $T = 10$, our alphabet includes 713 (x,y)-coordinates.} Then, the attack deploys a series of click-order heuristics to construct an attack dictionary. Our focus on these classes of attacks are motivated by their ease and minimal requirements for mounting by attackers, and their guessing ability  (especially for relatively large dictionary sizes).\footnote{Human-seeded attacks \citep{Thorpe2007} could have been alternatives for our analyses. However, those attacks usually have comparable guessing ability for relatively large dictionary sizes while requiring more information for mounting attacks, such as the actual background image, system error tolerance, collected sample password data on the same background image, etc. So this motivates us to focus on purely automated click-based attacks, which only require an image's dimensions and system error tolerance, which are fixed for a deployed PassPoints system.}   

There are three general classes of click-order based attacks. The \textit{LINE} class of attacks attempts to crack passwords that form a horizontal or vertical line across the background image. The \emph{DIAG} attack class guesses passwords with a dictionary of all possible straight lines, which are not necessarily vertical or horizontal. Note that for a fixed alphabet (i.e., fixed image and error tolerance), the DIAG dictionary always includes the LINE dictionary. The \emph{LOD} attack class attempts to crack passwords with a dictionary composed of passwords where each click-point is within a particular distance from its predecessor and successor. 

Each class of attacks has a \emph{relaxation parameter} $\tau$ (different from error tolerance $T$) controlling the extent to which each click-order pattern can be relaxed. The lower $\tau$ is, the more restrictive the pattern is. For example, LINE with $\tau=0$ generates those passwords exactly following a straight horizontal or vertical line, whereas LINE with $\tau=21$ allows two sequential click-points in a guessed password to deviate a maximum of $21$ pixels from a straight line. Letting $X(\tau)$ be the dictionary of an attack $X \in \{LINE, DIAG, LOD\}$ with relaxation parameter $\tau$, one can easily observe that, for $\tau'' > \tau'$, $X(\tau') \subseteq X(\tau'')$. In our experiments, we have used all three classes of attacks while varying the relaxation parameter $\tau$.

\addtolength{\tabcolsep}{+4pt}
\definecolor{Aqua}{rgb}{0.0, 0.7, 1.0}
\definecolor{orange}{rgb}{1.0, 0.55, 0.17}
\begin{table*}[tb]
    \begin{center}
    \addtolength{\leftskip}{-0.38cm}
        \begin{tabular}{|p{1.9cm}|p{0.9cm}|p{0.9cm}|p{0.9cm}|p{0.9cm}|p{1cm}|p{1cm}|p{0.9cm}|p{0.9cm}|p{0.9cm}|}
         \hline
         Group &LOD0 &LOD21 &LOD42 &LINE0 &LINE21 &LINE42 &DIAG0 &DIAG21 &DIAG42\\
         \hline
         \hline
         Grid:CTL&8.28\% &10.82\% &11.46\% &14.65\% &18.47\% &19.11\% &23.57\% &30.57\% &31.85\% \\
         Grid:RTL&\cellcolor{Aqua!25}5.26\% &8\cellcolor{Aqua!25}.42\% &\cellcolor{orange!25}11.58\% &\cellcolor{Aqua!25}12.63\% &\cellcolor{orange!25}21.05\% &\cellcolor{orange!25}22.11\% &\cellcolor{Aqua!25}17.89\% &\cellcolor{Aqua!25}27.37\% &\cellcolor{orange!25}32.63\%  \\
         Grid:LTR&\cellcolor{Aqua!25}6.4\% &\cellcolor{Aqua!25}6.4\% &\cellcolor{Aqua!25}8.8\% &\cellcolor{Aqua!25}11.2\% &\cellcolor{Aqua!25}16.0\% &\cellcolor{Aqua!25}18.4\% &\cellcolor{Aqua!25}20.0\% &\cellcolor{Aqua!25}27.2\% &\cellcolor{Aqua!25}28.8\%  \\
         \hline
         Highway:CTL &8.57\% &11.43\% &21.43\% &10.00\% &25.71\% &47.14\% &12.86\% &45.71\% &61.43\% \\
         Highway:RTL&\cellcolor{Aqua!25}0.0\% &\cellcolor{Aqua!25}2.0\% &\cellcolor{Aqua!25}4.0\% &\cellcolor{Aqua!25}0.0\% &\cellcolor{Aqua!25}10.0\% &\cellcolor{Aqua!25}28.0\% &\cellcolor{Aqua!25}8.0\% &\cellcolor{Aqua!25}30.0\% &\cellcolor{Aqua!25}44.0\% \\
         Highway:LTR&\cellcolor{Aqua!25}1.69\% &\cellcolor{Aqua!25}1.69\% &\cellcolor{Aqua!25}6.78\%  &\cellcolor{Aqua!25}1.69\% &\cellcolor{Aqua!25}13.56\% &\cellcolor{Aqua!25}28.81\% &\cellcolor{Aqua!25}11.86\% &\cellcolor{Aqua!25}32.2\% &\cellcolor{Aqua!25}49.15\% \\
         \hline
         Barn:CTL&8.86\% &10.13\% &10.13\% &11.39\% &13.92\% &18.99\% &12.66\% &26.58\% &31.65\% \\
         Barn:RTL&\cellcolor{Aqua!25}5.41\% &\cellcolor{Aqua!25}8.11\% &\cellcolor{orange!25}10.81\ &\cellcolor{orange!25}13.51\% &\cellcolor{orange!25}18.92\% &\cellcolor{Aqua!25}18.92\% &\cellcolor{orange!25}18.92\% &\cellcolor{orange!25}29.73\% &\cellcolor{orange!25}37.84\% \\
         Barn:LTR &\cellcolor{Aqua!25}0.0\% &\cellcolor{Aqua!25}2.63\% &\cellcolor{Aqua!25}2.63\%&\cellcolor{Aqua!25}0.0\% &\cellcolor{Aqua!25}2.63\% &\cellcolor{Aqua!25}2.63\% &\cellcolor{Aqua!25}2.63\% &\cellcolor{Aqua!25}10.53\% &\cellcolor{Aqua!25}15.79\%  \\
         \hline
         \end{tabular}
    \end{center}
    \caption{Percentage of passwords cracked with each attack class for various relaxation parameters $\tau$ = 0, 21, 42. The blue and orange colors encode settings with stronger and weaker security, respectively, compared to the corresponding control group's security.}
    \label{tab:LINE}
\end{table*}

Table \ref{tab:LINE} shows the percentage of guessed passwords for each group, image, and attacks. The relaxation parameter $\tau$ has been varied over \{$0$, $21$, $42$\} in our experiments. The LTR priming consistently enhances security (when compared to the control group) for all combinations of attacks and images. For RTL priming, in the majority of cases it enhances security, but the results are image-dependent. For Highway, the RTL priming exhibits higher password security than the control group for all attacks. However, for the two other images (Grid and Barn), the security outcomes for RTL priming vary depending on the attack. Next, we further analyse and discuss these results from various perspectives.

\vskip 2mm
\noindent \textbf{Grid Image.} For the Grid image, our primed groups exhibit slightly higher attack resistance compared to the control group on average (with a few exceptions for RTL). The improvement is relatively small for this image with the maximum improvements of 4.42\% for LOD21 and LTR.

\vskip 2mm
\noindent \textbf{Highway Image.} For Highway, both RTL and LTR consistently demonstrate considerable security improvements compared to the control group across all attacks. In LINE(0) and LOD(0), no passwords from the RTL group were successfully guessed at all. For other attacks, the passwords in the primed groups are 1.5-5  times more secure than those of the control group. Given that Highway is a compact-saliency image, where saliency is concentrated in a single region; see Figure~\ref{fig:bkg_img_sal}(d), priming might have been effective for improving security by counteracting the image's center bias. The priming techniques may lead users to select points outside of the highly salient center region, thus breaking up the simple geometric patterns that automated attacks target.

\vskip 2mm
\noindent \textbf{Barn Image.} The Barn image exhibits an interesting security pattern. For RTL, security is generally degraded, except for LINE(42), LOD(0), and LOD(21), where security is comparable or slightly improved. Surprisingly, LTR led to considerable security improvements for all classes of attacks. Its success might be attributed to exposing users to salient points on the left side of the image which they otherwise would not have been selected, thus breaking up the geometric patterns of passwords exploited by automated attacks.

\vskip 2mm
\noindent \textbf{Saliency and Security.} We intended to understand which type of saliency maps (compact vs diffuse) produces more secure passwords. Our results in Table \ref{tab:LINE} suggests that Barn as a representative of an image with a diffuse saliency map produces more secure passwords than others. In some cases, this security advantage is quite pronounced. For example, almost twice the percentage of passwords on the highway image's control group are cracked when compared to the that of the barn image for any attacks. This observation supports that the potential security of a background image is tightly linked to its underlying saliency. One can further observe that the saliency distribution of an image also limits the extent that priming can improve the security. For example, in the DIAG(42) attack on Highway and Barn, even though Highway's primed groups produced more secure passwords than the control group, they have still worse security than all groups on the Barn image. 

\begin{table}[tb]
    \centering
    \setlength{\tabcolsep}{11pt}
    \begin{tabular}{lccc}
        \toprule
         & \textbf{Grid} & \textbf{Highway} & \textbf{Barn} \\
        \midrule
        \textbf{Control} & 31.85\% & 61.43\%  & 31.65\%\\
        \textbf{Primed} &  30.45\% & 46.79\%  & 26.67\%\\
        \midrule
        Improvement & 1.40\%& 14.64\% & 4.98\%\\
        \bottomrule
    \end{tabular}
    \caption{Percentage of Passwords Cracked by DIAG(42) for Primed (i.e. LTR and RTL) and Control groups}
    \label{tab:expected_priming}
\end{table}

\vskip 2mm
\noindent \textbf{Does priming enhance security?} To answer this question, we draw our attention to the most powerful attack of DIAG(42), which is the superset of all other attacks. As the users are randomly assigned to RTL or LTR, we can merge these groups to the \textit{Primed} group, and then attack the merged set with DIAG(42). As shown in Table \ref{tab:expected_priming}, we can observe that priming has enhanced security. However, the extent of improvement is image dependent ranging from 1.40\% for Grid to 14.64\% for Highway.

\subsection{Usability}
We collected user-reported usability data from the 107 users who completed all three sessions of the study for the Grid image. We examine this data for all three groups (RTL, LTR, and Control) to determine if the priming techniques impact the perceived usability of the system. For these comparisons, we combine the RTL and LTR groups into a single \textit{Primed} group, and compare their usability results to the control group. We also compare login times for each session, and memorability metrics such as the number of password resets and incorrect login attempts. 

The exit survey asks users to report if they recorded their passwords during the study. Our implementation attempts to mitigate password recording with two approaches; never displaying a user's password to them, in part or in full, and providing no visual feedback as to the location of click-points during creation. Among 107 users, 5 (12.8\%) of 44 users in the Control group and 2 (3.2\%) of 63 users in the Primed group reported recording their passwords. This suggests that password recording is no more prevalent in graphical passwords than in traditional text passwords, with password recording rates of 17-50\% \citep{Komanduri2011}. 
For our usability analyses, we exclude 5 users from the Control group and 2 users from the Primed group who reported recording their passwords. We therefore have 39 users in our Control group and 61 users in our Primed group. 

\subsubsection{System Usability Scale}
To compare usability between these groups, we use the System Usability Scale (SUS) \citep{brooke1996}. The SUS is a survey with 10 short Likert Scale questions, and asks the user to score them from one to five, where one is strongly disagree and five is strongly agree. The tone of the questions alternates: for odd-numbered questions, a score of 5 is an indicator of good usability whereas for even-numbered questions, 5 is an indicator of bad usability. This tone alternation discourages users from answering all questions with 1 or 5, and aids in detecting careless all-one or all-five answers to the questions. The SUS score for each user is calculated by subtracting 1 from each odd-numbered question response, subtracting each even-numbered question response from 5, summing all ten resulting values, and then multiplying by 2.5. This yields a score between 0 and 100 for each user. Scores above 68 indicate above average usability.

The Control group and Primed group had an average SUS score of 74.68 (SD=16.005) and  77.30 (SD=17.481), respectively. We conduct a two-sided independent t-test between the SUS scores of the Control and treatment groups where the independent variable is the group, the dependent variable is the SUS score, with 98 degrees of freedom. This test fails to reject the null hypothesis that both samples come from the same distribution, $t(98)=-0.7603$, $p=0.4491$. 
This suggests that the priming techniques do not have a negative impact on user perception of usability.

Users in treatment groups were also asked if they noticed that the priming effect was present during creation, but not login time. Users who answered yes were asked to rate the following Likert Scale statement on a scale of one (strongly disagree) to five (strongly agree): ``I found the image revealing annoying.'' Of 61 users, 23 (37.7\%) users reported noticing the effect during creation, but not during logins. Of these 23 users, 8 (34.7\%) of the users agreed or strongly agreed with the statement, while 9 (39.1\%) of the users disagreed or strongly disagreed and 6 (26.1\%) of the users responded neutrally.

\subsubsection{Login Time}
We aim to study the users' login times in the Primed and control groups. A user's login time starts from when the image is displayed until the successful login (i.e., the time for successful login). For users who reset their password during the study, login time is recorded for their latest password. For each session, we perform a two-sided independent t-test where the independent variable is the user's group (e.g., Control or Primed) and the dependent variable is the login time measured in seconds with 98 degrees of freedom.

For Session 1, the average login time for users is 21.342 seconds (SD=16.126) and  21.476 seconds (SD=13.947) in the Control and Primed groups, respectively. We fail to reject the null hypothesis that both groups have the same average login time, $t(98)=-0.0439$, $p=0.9651$. 
For Session 2, the login times for Control ($27.004\pm22.008$) and Primed ($28.774 \pm 20.985$) are not statistically different, $t(98)= -0.3997$, $p=0.6903$. 
Similarly, the login times for Control group ($26.158\pm19.458$) and Primed group ($29.975 \pm 26.163$) are not statistically different in Session 3, $t(98)= -0.7753$, $p=0.4400$. 
These analyses allow us to conclude that that there is no statistically significant difference in login times between Primed and Control users for all three sessions.

\subsubsection{Memorability}
We compare password memorability between the Primed and Control groups of the Grid image by considering the number of password resets and unsuccessful login attempts for each group. While previous work has applied the \emph{memory time} \cite{Stobert-memory-time} metric to evaluate the memorability of passwords, we believe our low number of password resets
and the fixed time window of our login sessions limit the additional information this metric provides for our data. 

\vskip 10pt
\noindent \textit{Password Resets.} We count the number of users with password resets in both the Primed and Control groups across all sessions. No users in the Control group reset their password for any session. One user in the Primed group reset their password during Session 2.  We compare the two groups using Fisher's exact test where the independent variable is the user group, the dependant variables are the frequency of resetting and non-resetting users in each group, with 1 degree of freedom. 
We find no significant difference in the frequency of resets between the two groups, $p=0.9999$. 
This suggests that the priming effect did not cause users to forget their passwords more often than the Control group.

\vskip 10pt
\noindent \textit{Unsuccessful Login Attempts.} 
For all sessions, we record the number of unsuccessful login attempts. 
We compare the average number of unsuccessful login attempts between the Primed and Control groups with a two-sided Mann-Whitney-U test. 
We fail to reject the null hypothesis that the frequency of password resets in either group is the same for Session 1 ($U=1179.0$, $p=0.9442$), Session 2 ($U=1080.0$, $p=0.4413$), and Session 3 ($U=1181.5$, $p=.96012$).  
Again, this suggests that the priming effects do not have a significant effect on the memorability of a selected password.

\section{Limitations}
Although the use of MTurk comes with many advantages, it might have some limitations. 
While studies performed on MTurk can have comparable results to those performed in lab settings \citep{crump2013}, users might have low incentive to perform as well as a natural setting.  To minimize any such impact on measurements of the priming effects, we filter our data based on Session 1 questionnaire responses as described in Section \ref{sec:filtering}. Even if this filtering missed some unmotivated participants, our results would simulate a worst-case scenario wherein users do not pay attention to the image presentations used in our study, in which case our results under-report the effect of the image presentations we study. 

Our image selection process is semi-automated, and required one of the researchers to manually review the clusters produced by each possible set of features. As this review was conducted by a single researcher, there is a potential for subjective bias to be introduced around what constitutes a \textit{correct} clustering of the images.

For the Grid image, we retained 42\% of users by the end of Session 3 before filtering out users who were distracted during the image reveal---a retention rate consistent with similar studies performed on MTurk (e.g., 55\% \citep{Shay12} and 42-51\% \citep{Joudaki18}). This drop-off may have inflated the usability results, if users who disliked the system chose to drop out. However, such inflation should apply equally to all groups, so it should not impact comparisons. 

The duration of our study was for approximately one week (chosen to approximate the login frequency of online bank accounts \cite{Hayashi2011}), therefore our memorability findings should not be considered for authentication systems with long delays between logins (e.g., fallback/secondary authentication). However, the duration of the study does not affect the results on password choice, which is the main goal of image presentations in this work. 

The exit survey in Session 1 asked about user's security skill. This might have hinted some users that the study involves security evaluations, which consequently may have influenced their behavior. However, as the survey was given after password creation, only users who reset their passwords would have had this information at the password creation time. 

Our system provides no visual feedback during password creation to avoid users to easily record (e.g., take a picture of) their passwords. While this decision was justified to mitigate recording, it has the potential to decrease the memorability of the password by not allowing users to visualize their password as a whole.  As a result, our memorability results may not be comparable to other systems where visual feedback is provided. 
Thus, we focus on the comparison of memorability between user groups, all of which were exposed to this design choice, rather than comparing memorability in absolute terms with other studies.

The majority of our participants were under the age of 35. While this is consistent with the demographics of previous graphical password users studies, and is less skewed than studies with only university student participants, it may present generalizability issues. Age related differences in processing visual stimuli and locating salient points may impact a user's password selections. Our results might therefore not be applicable to older age groups.
\begin{figure*}[t]
    \begin{center}
        \begin{tabular}{ccc}
            \includegraphics[width=0.30\textwidth]{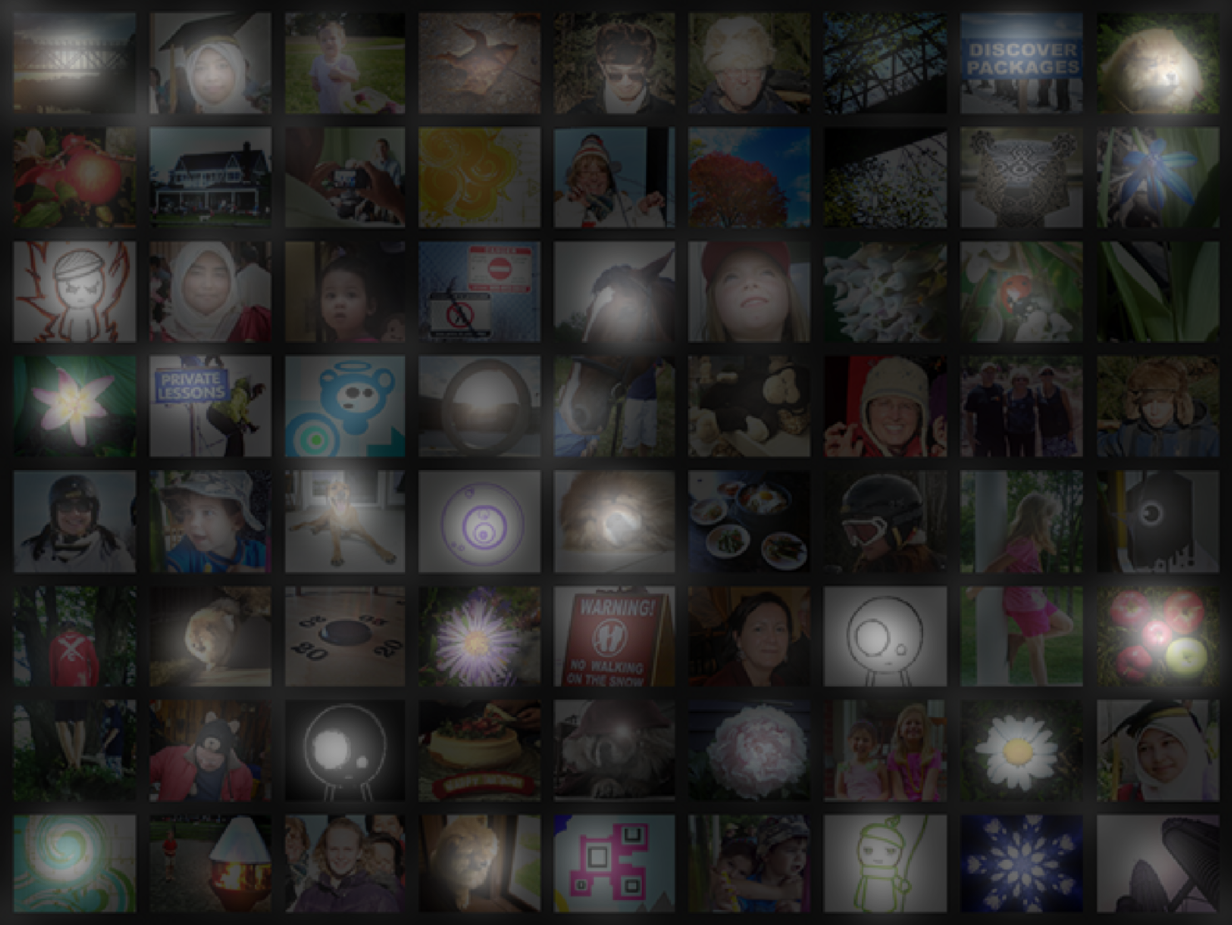} &
            \includegraphics[width=0.30\textwidth]{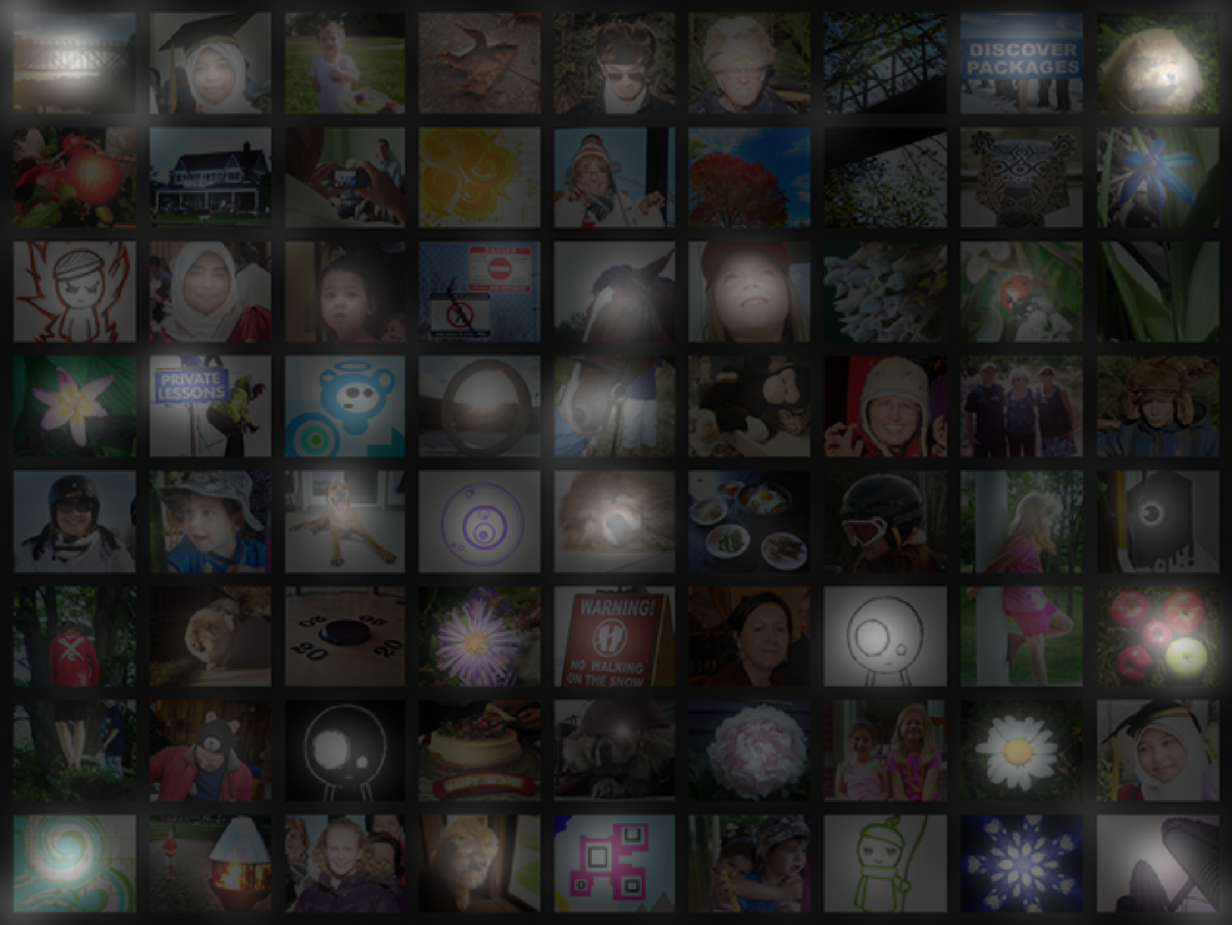} &
            \includegraphics[width=0.30\textwidth]{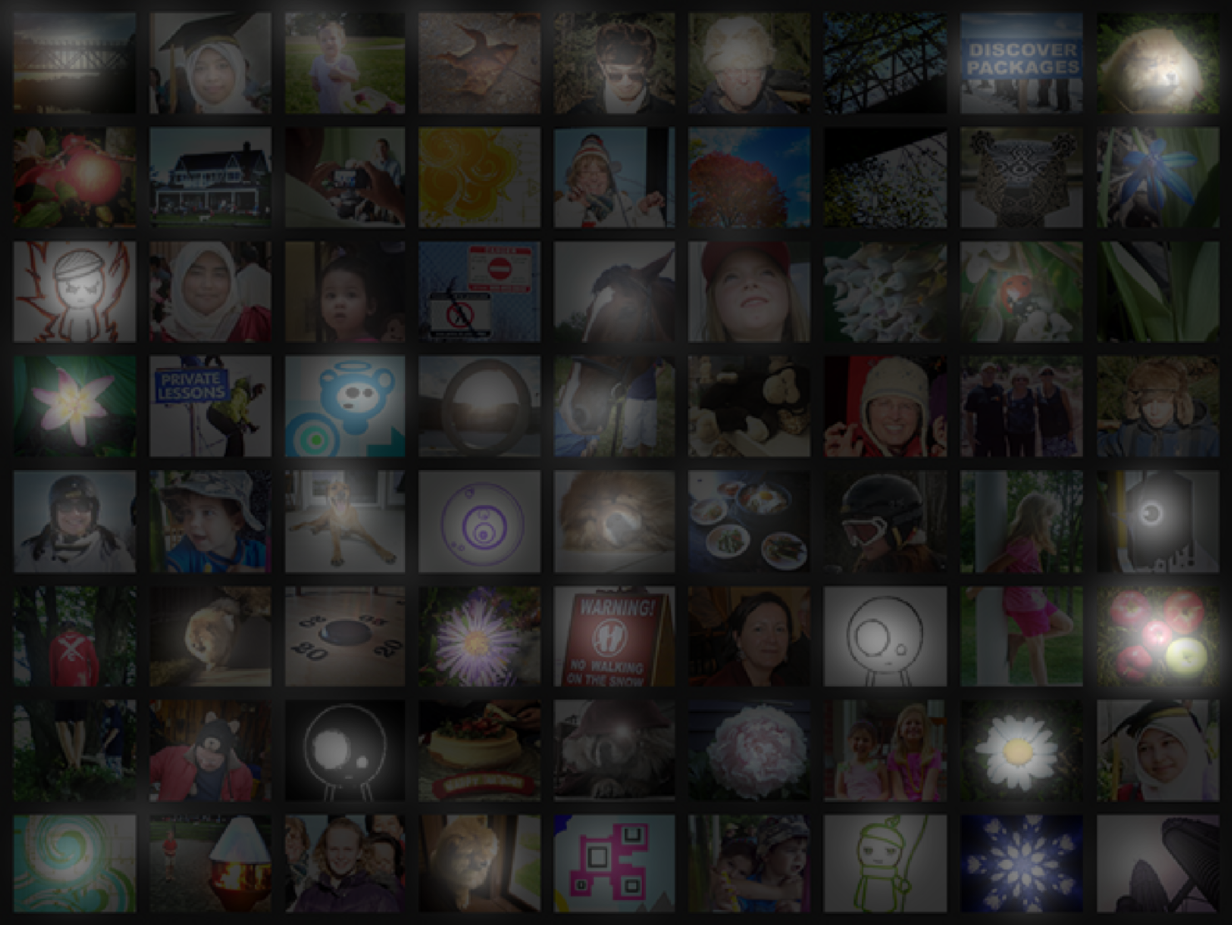} \\
            (a) Grid Image LTR & (b) Grid Image Control & (c) Grid Image RTL\\
            \includegraphics[width=0.30\textwidth]{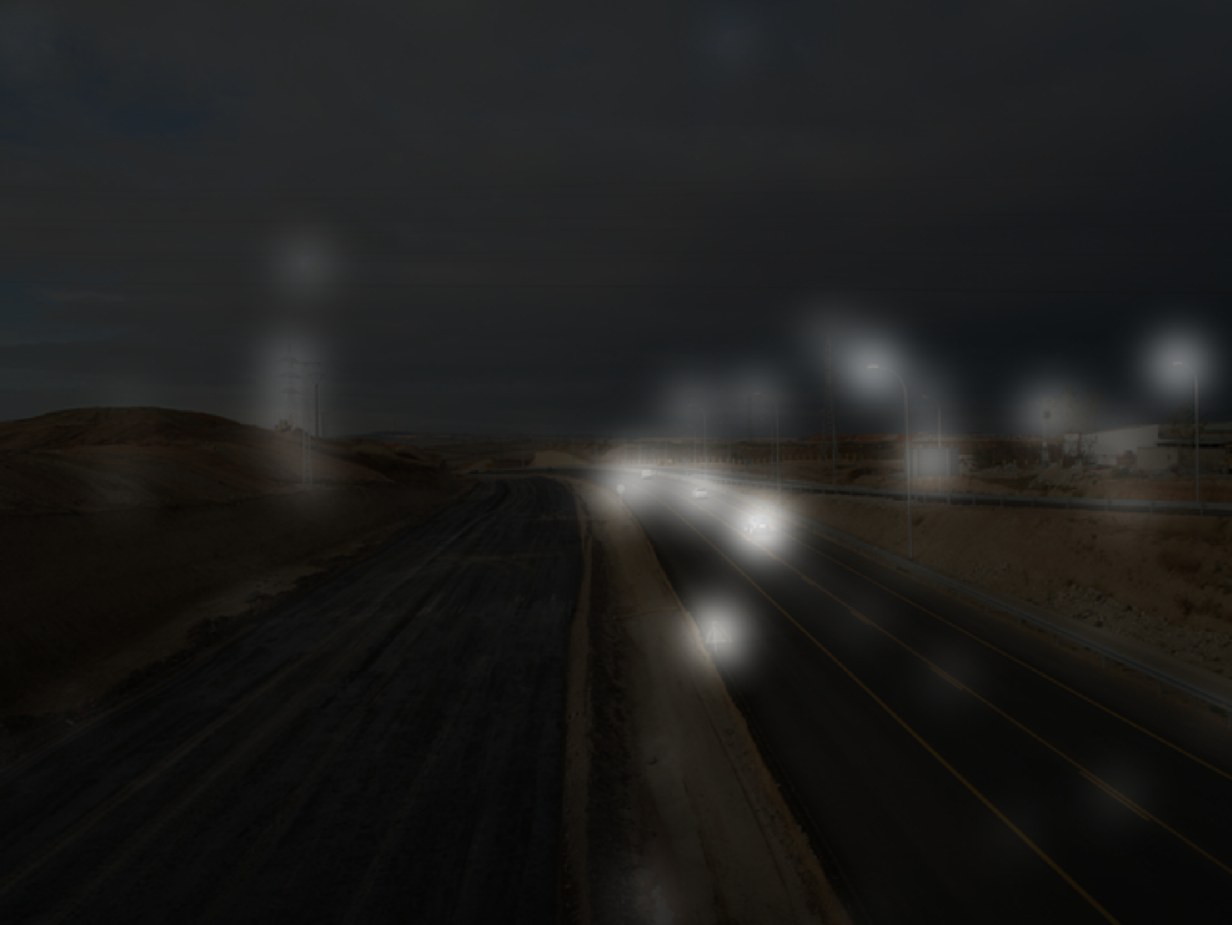} &
            \includegraphics[width=0.30\textwidth]{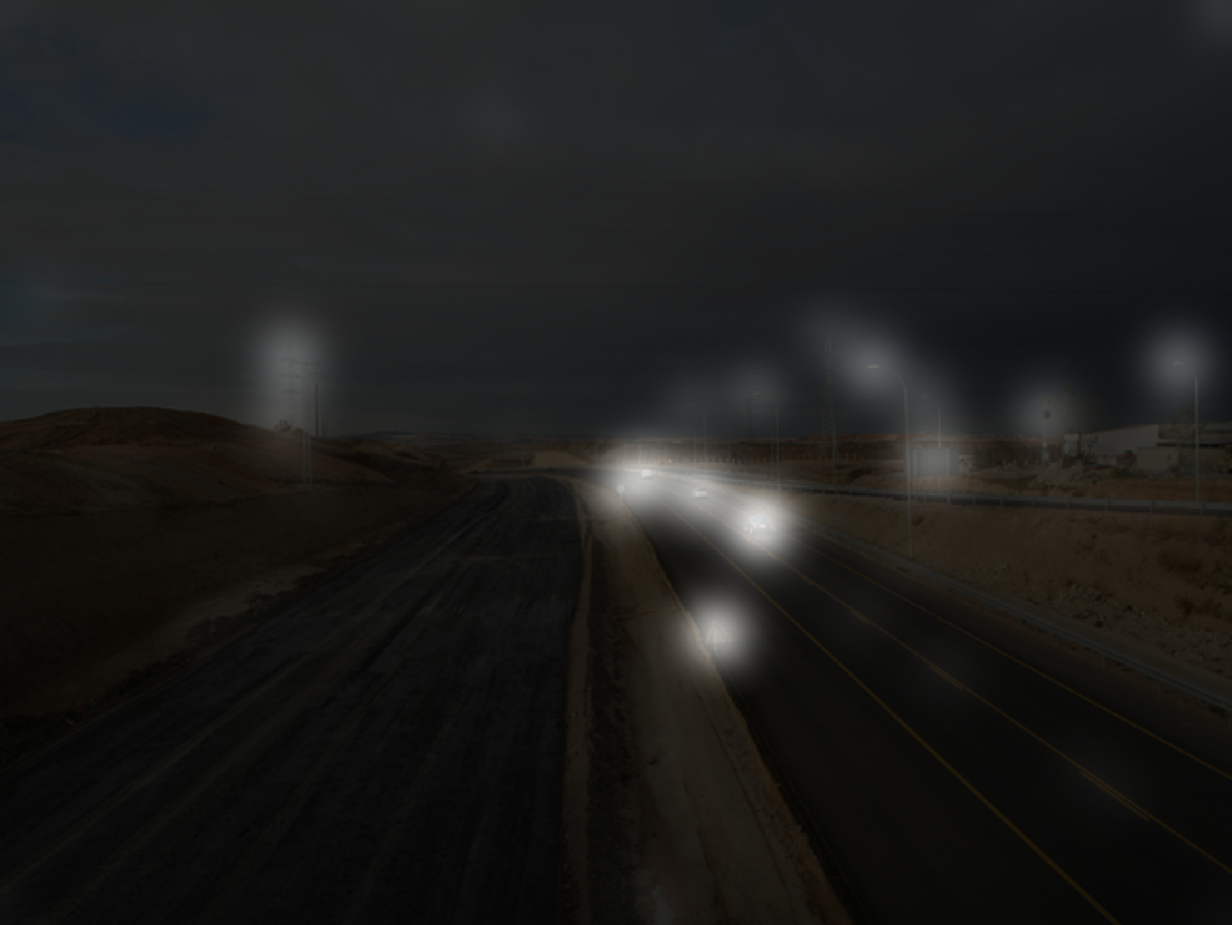} &
            \includegraphics[width=0.30\textwidth]{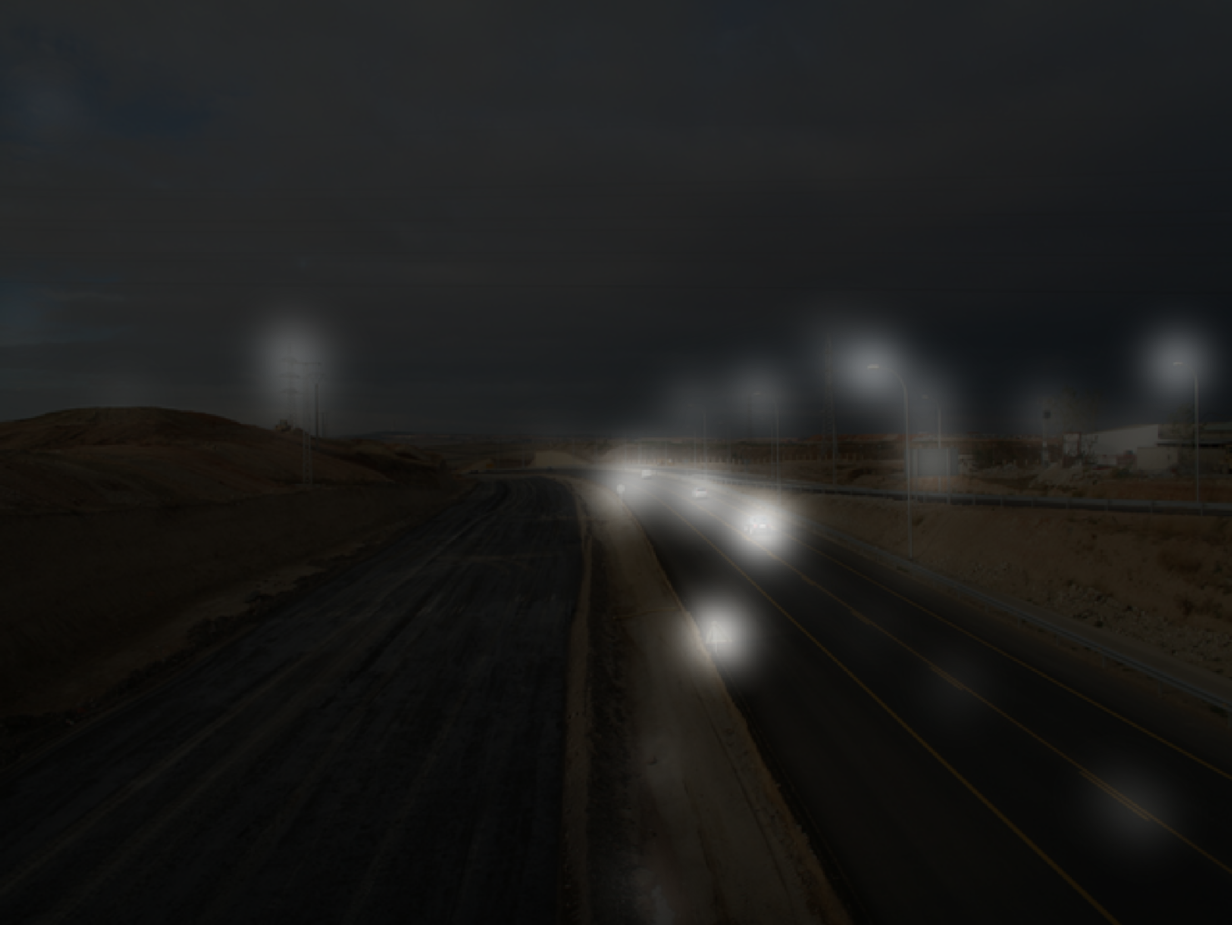} \\
            (d) Highway Image LTR & (e) Highway Image Control & (f) Highway Image RTL \\
            \includegraphics[width=0.30\textwidth]{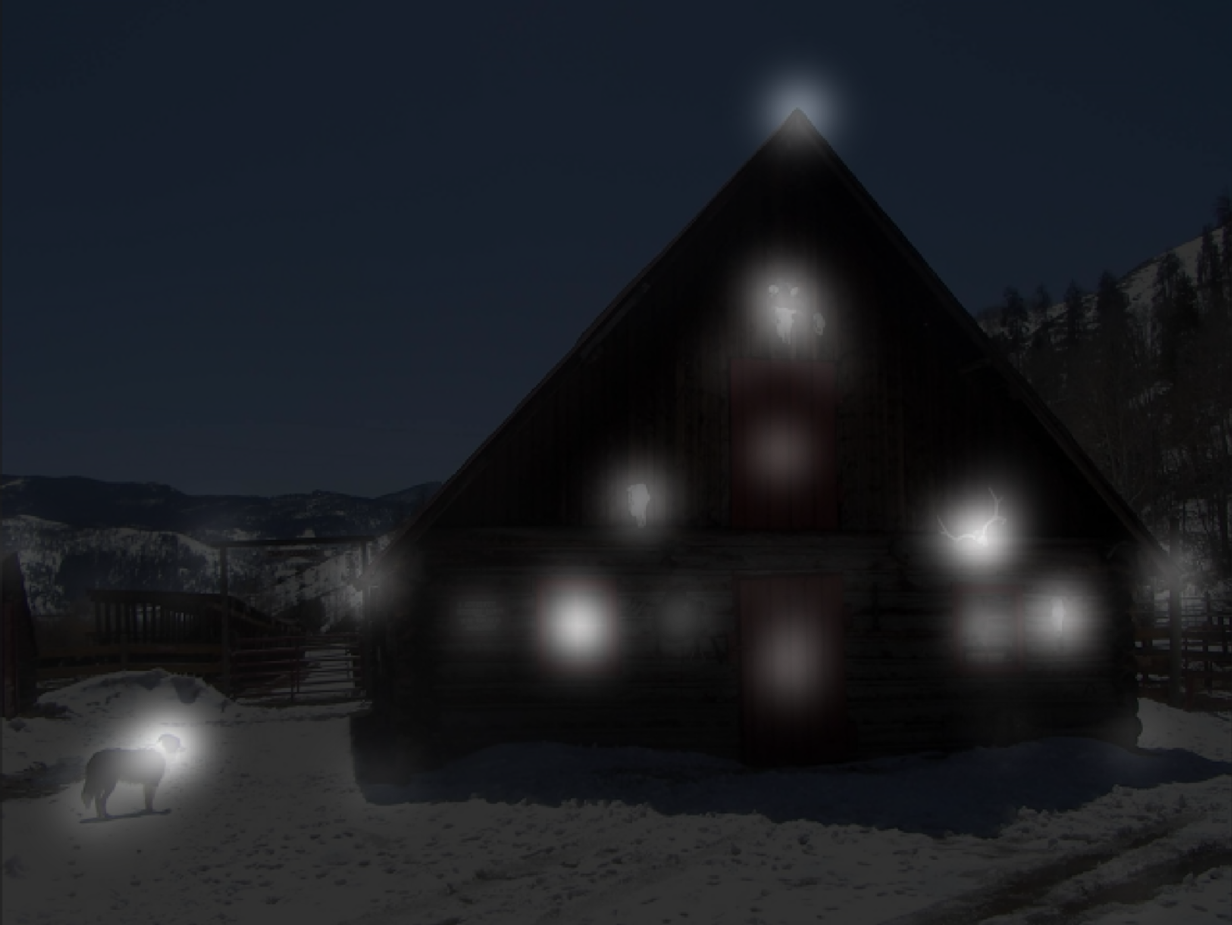} &
            \includegraphics[width=0.30\textwidth]{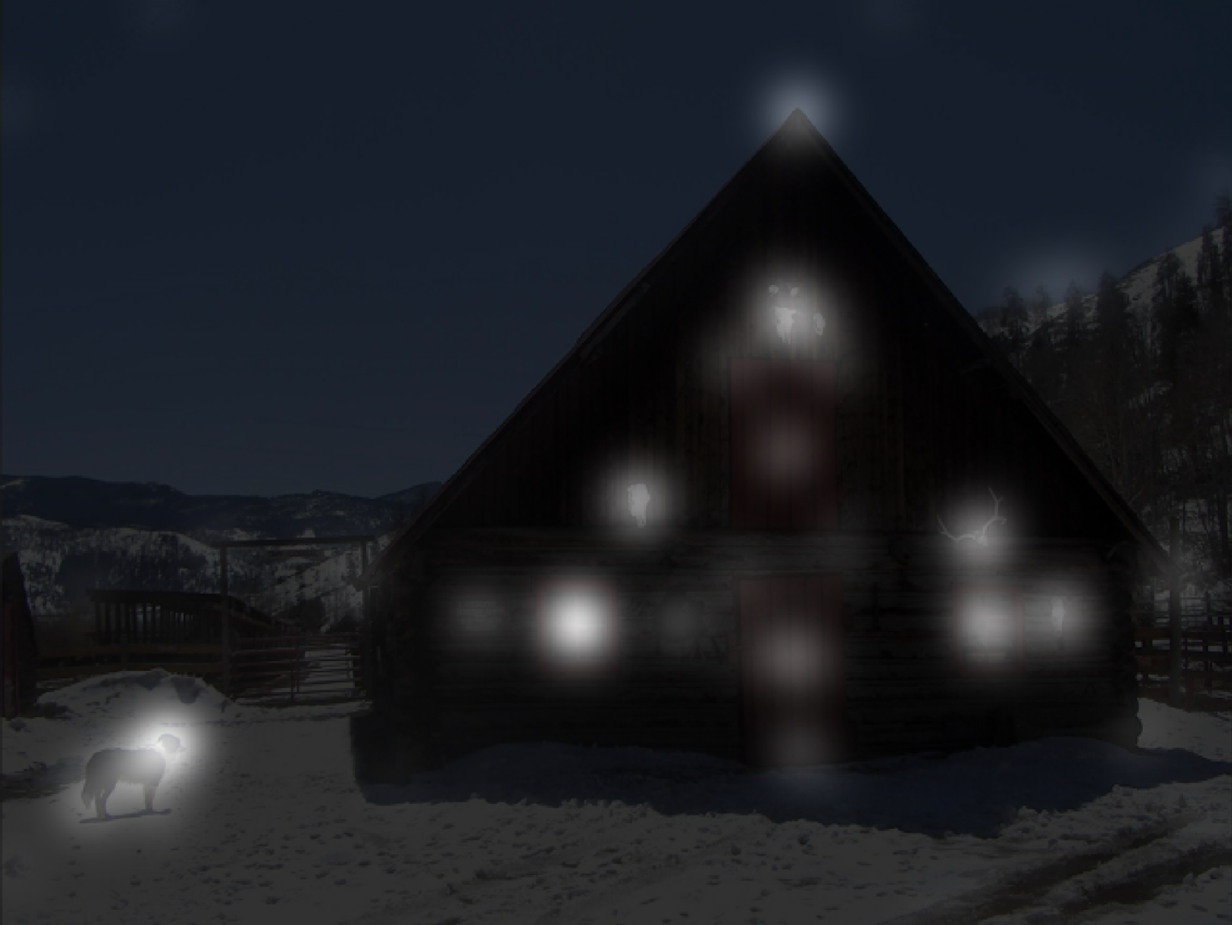} &
            \includegraphics[width=0.30\textwidth]{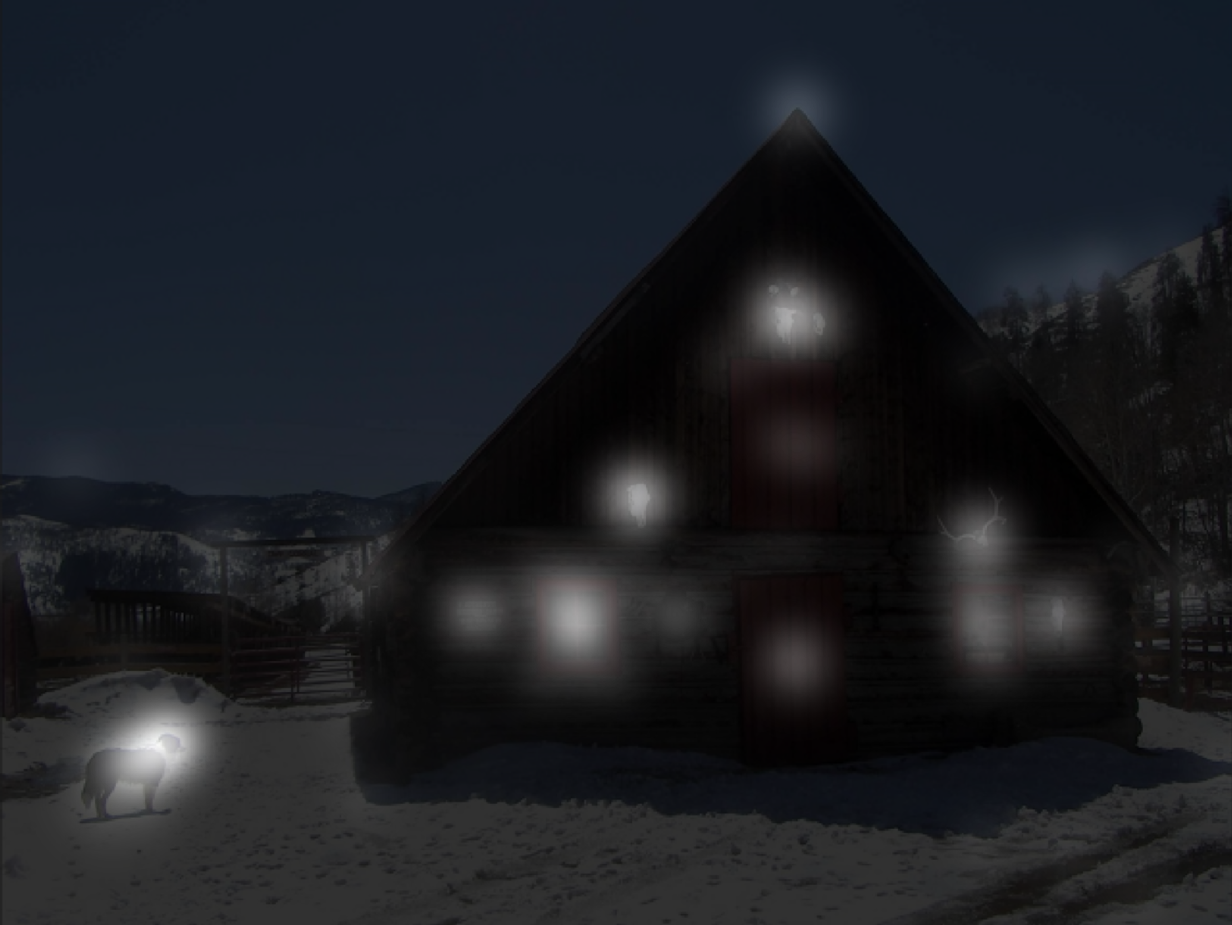} \\
            (g) Barn Image LTR & (h) Barn Image Control & (i) Barn Image RTL\\
        \end{tabular}
    \end{center}
    \caption{Heat maps generated from the click-points of each group when superimposed onto the background image used during password creation.}
\label{fig:bkg_img_heat}
\end{figure*}

\section{Discussion}
In terms of usability, our findings indicate that \emph{drawing the curtain} has no negative impact. There were no statistically significant differences between the overall perception of the system between the control and treatment groups (as measured by SUS scores). Additionally, the examined priming methods didn't have any significant impact on the login times or memorability as observed through password resets and unsuccessful login attempts. However, 34.7\% of users found the drawing-the-curtain priming technique (during the password creation) to be annoying. 

In terms of security, the click-order attack results revealed a complex interplay between priming techniques employed, background images, and the resulting graphical passwords that users choose.  In particular, primed users for our image with a compact saliency map (Highway) fared much better against attacks than the control group. The primed groups had mixed results for the image with diffuse saliency map (Barn)---RTL fared worse, but LTR fared better than the control group. These results suggest that the presentation effect is indeed image dependent, and furthermore only some image presentations can influence an effect on some images. It could be the case that images with compact saliency information present more opportunity for improvement. The concentration of saliency in a single region may lead users to select points within a smaller region, leading to the worse security outcomes. The priming effect can bias some selections outside of this region, and consequently increase the password security. In contrast, for the image with the diffuse saliency map, the salient points are already dispersed enough so certain directional priming techniques produce less noticeable changes to password choices. 

In terms of the impact of the image presentations on individual click-point selections, when we compared each primed group against the control group, we found no significant changes in the click-point distribution with a Mann-Whitney-U test (for all of the three images studied).
When we compare the click-points (for all 5 selections) of the RTL and LTR groups for each image (Figure \ref{fig:bkg_img_heat}), we notice that for the Highway and Barn images, the click-point locations of the primed groups do not vary greatly from the control group. While different locations are selected, or selected more frequently across the passwords for those groups, the overall effect is subtle, as demonstrated by the low brightness of the hot-spots on Fig \ref{fig:bkg_img_heat}. This suggests that changes in security, against click-order attacks, might be at least partly the result of changes in the ordering of click-points rather than altering hot-spot locations. 


The most common self-declared password selection strategies are attention and geometric strategies across all images. For Highway and Barn, attention-based methods were most common (43.6\% and 32.5\% respectively). For the Grid image, geometric strategies were the most common (34.0\%), which is not surprising given the image's structure. 

In interpreting the sum of these results, it is important to consider that the purpose of the statistical tests is to detect that the image presentations have an impact on user choice of individual click-points. The statistical tests simply test to see whether, for each click taken in isolation, the distribution has changed. However, the click-order attacks reveal changes to the percentage of easily guessed entire passwords (the ordered set of 5 click-points). This is arguably a more meaningful test of a change in behaviour, especially in relation to security outcomes.  

We stress that our results for \emph{drawing the curtain} do not necessarily apply to other priming techniques. For example, Katsini et al.\ \citep{katsini2018} found that image presentations based on saliency maps produced stronger passwords, especially when the image presentations were tailored to the user's cognitive style. 

Without a controlled lab environment, it is difficult to know whether users have actually viewed the image presentation. As such, our results may underestimate the impact of image presentations. 
In general, identifying methods to ensure user engagement during priming will be important for the future success of priming-based approaches.

\section{Concluding Remarks and Future Directions}
We performed a large-scale study on drawing-the-curtain priming techniques on PassPoints.  Our findings include: (i) These priming techniques do not impact usability, and (ii) There are security benefits offered by the priming techniques employed, but these security benefits are dependent on both the background image and priming technique used. The results indicate that these priming techniques need to be carefully designed.

We found that drawing-the-curtain priming effects can improve the security of passwords selected on background images with highly concentrated saliency (i.e., in the compact cluster). Future work could extend this analysis by observing the security effects of the priming technique over images from a more diverse range of categories. 
This may assist in increasing the number and types of images that can be safely used for graphical password backgrounds. Future work should also seek to develop and test different image priming techniques. In the realm of curtain drawing, top-to-bottom, bottom-to-top, and oblique curtain draws offer clear extensions to this class of priming methods.

Future work might also focus on determining a method to automatically select or produce priming techniques that are tailored to a specific background image, and consider all possible security implications of the image presentation (i.e., hot-spots in the click-point distribution and click-order patterns). Examining structural properties of background images as well as saliency maps, especially those generated by saliency predictors, may be a useful starting point.

Automatically generated priming techniques should also be carefully constructed in order to resist attackers with knowledge of both the image and the applied priming technique. There are two general approaches to resist such attacks.  The first is to apply priming such that it increases the entropy of the password space, rather than simply biasing users to select points in a different ordering or location. The second is to apply a non-deterministic priming technique that is selected from a large pool of candidates.

From a usability perspective, one might seek to find classes of users who find particular priming techniques to be unobtrusive. In this way, the selection of a priming technique could be both based on the background image, and personalized to mitigate the technique's ability to degrade user experience.

\bibliographystyle{ACM-Reference-Format}
\bibliography{mybibfile.bib}

\appendix

\end{document}